\documentclass[prd,showpacs,amsmath,amssymb,floatfix,twocolumn]{revtex4}

\usepackage{bm}
\usepackage{dcolumn}
\usepackage{color,graphicx}
\usepackage{feynmp}

\renewcommand{\vec}[1]{\bm{#1}}
\newcommand{\grad}{\bm{\nabla}}
\newcommand{\rhobar}{\bar{\rho}}
\newcommand{\pmat}[1]{\begin{pmatrix} #1 \end{pmatrix}}

\newcommand{\bx}{\vec{x}}
\newcommand{\bk}{\vec{k}}
\newcommand{\bq}{\vec{q}}
\newcommand{\bL}{\vec{L}}
\newcommand{\bP}{\vec{\Psi}}

\begin{document}
\unitlength = 1mm

\begin{fmffile}{diagrams}
\fmfwizard
\fmfcmd{%
    vardef ocross =
        for i = 1 upto 4:
            origin --
            for j = 0 upto 10:
                (0.5up rotated (90i-45+90j/10)) --
            endfor
        endfor
        cycle
    enddef;
}
\newcommand{\fmfbox}[4]{\parbox{#1}{\fmfframe(#3)(#3){\begin{fmfgraph*}(#2)#4\end{fmfgraph*}}}}
\newcommand{\fmfo}[1]{\fmfv{decor.shape=circle,decor.filled=empty,decor.size=3mm}{#1}}
\newcommand{\fmfocross}[1]{\fmfv{decor.shape=ocross,decor.filled=empty,decor.size=3mm}{#1}}

\title{A Critical Look at Cosmological Perturbation Theory Techniques}

\author{Jordan Carlson}
\email{jwgcarlson@berkeley.edu}
\affiliation{Department of Physics, 366 LeConte Hall, 
University of California Berkeley, CA 94720}

\author{Martin White}
\email{mwhite@berkeley.edu}
\affiliation{Departments of Physics and Astronomy, 601 Campbell Hall,
University of California Berkeley, CA 94720}

\author{Nikhil Padmanabhan}
\email{NPadmanabhan@lbl.gov}
\affiliation{Physics Division, Lawrence Berkeley National Laboratory,
1 Cyclotron Rd., Berkeley, CA 94720}

\begin{abstract}
Recently a number of analytic prescriptions for computing the non-linear matter
power spectrum have appeared in the literature.  These typically involve
resummation or closure prescriptions which do not have a rigorous error
control, thus they must be compared with numerical simulations to assess
their range of validity.  We present a direct side-by-side comparison of
several of these analytic approaches, using a suite of high-resolution N-body
simulations as a reference, and discuss some general trends.
All of the analytic results correctly predict the behavior of the power
spectrum at the onset of non-linearity, and improve upon a pure linear
theory description at very large scales.  All of these theories fail at
sufficiently small scales.  At low redshift the dynamic range in scale where
perturbation theory is both relevant and reliable can be quite small.
We also compute for the first time the 2-loop contribution to standard
perturbation theory for CDM models, finding improved agreement with
simulations at large redshift.
At low redshifts however the 2-loop term is larger than the 1-loop term on
quasi-linear scales, indicating a breakdown of the perturbation expansion.
Finally, we comment on possible implications of our results for future studies.
\end{abstract}


\maketitle

\section{Introduction}

The character and evolution of the large-scale structure of the Universe has
been the subject of much research in recent decades.  As it is currently
understood, large-scale structure grows through a process of gravitational
instability starting from a nearly scale-invariant spectrum of Gaussian
fluctuations at early times.  On very large scales the matter distribution of
our universe today is well modeled by linear perturbation theory.  On scales
below about $10\,$Mpc, on the other hand, the dynamics are highly non-linear
and we must resort to direct numerical simulations of the N-body problem to
understand the clustering of matter or its tracers.

On intermediate, or quasi-linear, scales there is the possibility that the
matter distribution may be modeled analytically by extending perturbation
theory beyond linear order.  This possibility has received renewed attention
recently due to the interest in using baryon acoustic oscillations as a probe
of the expansion history of the Universe and of the nature of dark energy
\cite{Eisenstein05}.  Since the baryonic features are at large scales
($\mathcal{O}(100)\,$Mpc) it is plausible that higher order perturbation theory
could model subtle corrections to the linear result with some accuracy.  More
generally, investigation of perturbation theory may allow some improvement in
theoretical predictions for the next generation of very large surveys.

Consequently, a number of new ideas have been introduced in recent years for
computing statistical properties of the matter distribution, most importantly
the 2-point function or power spectrum.  Regrettably these approaches involve
uncontrolled approximations, providing no simple way of estimating the
theoretical uncertainty.  Since perturbation theory is expected to fail on
sufficiently small scales, the domain of validity of any particular approach is
therefore unclear, and the only known way to test their accuracy is to compare
their predictions with the results of N-body simulations.
In the past this has been done on a case-by-case basis, with one theory tested
for one cosmology against one suite of N-body simulations, focusing primarily
on the power spectrum.  Recently there have been some attempts to compare
multiple theories simultaneously \cite{Nishimichi08,Taruya08}, or to examine
statistics other than the power spectrum
\cite{Bernardeau08,Valageas08,Guo09}.
However a comprehensive comparison has been lacking, and with the recent
proliferation of analytic techniques it is natural to ask how well these
theories actually perform.  With near-future observations potentially depending
on these techniques and with recent advances in N-body algorithms and computing
power, it is timely to revisit this issue.

In this paper we present a direct comparison of several recent analytic
predictions for the clustering of matter on quasi-linear scales.  We restrict
our attention to the matter fluctuations, because very few of the existing
treatments can handle biased tracers such as dark matter halos and galaxies.
We use modern, high-resolution N-body simulations as our reference points,
which provide highly accurate \cite{Evrard08,Heitmann08,CoyoteI} (though
computationally expensive) estimates for statistical observables of the matter
distribution.  By comparing the analytic predictions for two cosmologies,
one close to the current best-fit model and one more extreme, we are able
to judge the relative merits of each approach.

The paper is organized as follows.  In Section \ref{sec:dynamics} we start by
reviewing the dynamical equations that govern the evolution of the matter
distribution and discuss the relevant statistical quantities that one may
compute.   We then continue by summarizing the different analytic approaches we
consider in this paper.  In Section \ref{sec:sims} we describe the N-body
simulations that are used as a reference point for the comparison.  In Section
\ref{sec:compare} we plot the various approaches together, discuss
qualitatively how well they agree with simulations, and propose several ways to
quantify this agreement.  We discuss the results of this comparison in Section
\ref{sec:discuss}, and make some closing remarks in Section \ref{sec:conclude}.

\section{Analytic Methods}
\label{sec:dynamics}

We start by reviewing the different analytic methods we consider - our goal is 
not to provide a comprehensive description of each method, but to provide an
overview and highlight the relationships between the different methods.  

\subsection{Dynamics and Linear Theory}

By far the most popular approach to an analytic description of large-scale
structure is to approximate the matter distribution as an irrotational fluid,
characterized by a density constrast
$\delta(\vec{x}) = \rho(\vec{x})/\rhobar - 1$ and a
peculiar velocity divergence $\theta(\vec{x}) = \grad\cdot\vec{v}(\vec{x})$.
The fluid equations, in Fourier space, are then (see
Appendix~\ref{sec:eptreview} for a detailed derivation),
\begin{widetext}
    \begin{subequations}
        \label{eq:fluidk}
    \begin{align}
        \frac{\partial\delta(\vec{k})}{\partial\tau} + \theta(\vec{k})
            &= -\int \frac{d^3q}{(2\pi)^3} \frac{\vec{k}\cdot\vec{q}}{q^2}
                \,\theta(\vec{q})\delta(\vec{k}-\vec{q}), \\
        \frac{\partial\theta(\vec{k})}{\partial\tau} + \mathcal{H}\theta(\vec{k})
                + \frac{3}{2}\Omega_m \mathcal{H}^2 \delta(\vec{k})
            &= -\int \frac{d^3q}{(2\pi)^3}
                \frac{k^2 \,\vec{q}\cdot(\vec{k}-\vec{q})}{2q^2 |\vec{k}-\vec{q}|^2}
                \,\theta(\vec{q})\theta(\vec{k}-\vec{q}).
    \end{align}
    \end{subequations}
\end{widetext}
Here $d\tau = dt/a(t)$ is conformal time, $\Omega_m(\tau) =
\rhobar(\tau)/\rho_\text{crit}(\tau)$, and $\mathcal{H} = aH$ is the conformal
Hubble parameter.  (Note that we adopt the Fourier transform convention that
puts the $(2\pi)^3$ in the wavevector integral.  We also omit the tilde that is
usually used to decorate Fourier space quantities.)  The non-linear nature of these 
equations is manifest in the mode-coupling integrals.

Working to linear order in $\delta$ and $\theta$, we obtain 
\begin{equation}
    \delta_L(\vec{k};z) = \frac{D(z)}{D(z_i)} \delta_i(\vec{k}) \,\,
\end{equation}
and 
\begin{equation}
    \theta_L(\vec{k};z) = -\mathcal{H}(z)f(z)\frac{D(z)}{D(z_i)}\delta_i(\vec{k}),
\end{equation}
where $\delta_i$ is the density contrast at some early time $z_i$ when linear
theory is certainly valid, $D$ is the linear growth function (normalized to 1
at $z = 0$), and $f \equiv d\ln D/d\ln a$.  At early times $\Omega_m \approx 1$
and $D \propto a$.  For convenience we define $\delta_0$ to be the linear
density contrast today, i.e. $\delta_0(\vec{k}) = \delta_L(\vec{k};z=0)$.  When
convenient we also follow common practice and use $\eta = \ln D$ as a time
variable; for brevity we often suppress the time dependence of quantities
altogether. It is further convenient to group $\delta$ and $\theta$ into a
2-component vector,
$\Phi_a(\vec{k})=(\delta(\vec{k}),-\theta(\vec{k})/\mathcal{H}f)$ which is
proportional to $(1,1)$ in linear theory.

\subsection{Statistical observables}

Inflation predicts, and observations have confirmed, that the initial
fluctuations are predominantly adiabatic \cite{WMAP}, almost scale-invariant
\cite{WMAP}, and very close to Gaussian \cite{Slosar08}.  Under the assumption
that the initial field is Gaussian all expectation values of moments of the
evolved density and velocity fields can be expressed as integrals over the
linear theory power spectrum.  For example, the evolved 2-point function
\begin{equation}
\label{eq:pk}
    (2\pi)^3 \delta_D(\vec{k}+\vec{k}') P_{ab}(k)
        = \left\langle \Phi_a(\vec{k}) \Phi_b(\vec{k}') \right\rangle,
\end{equation}
whose components are all equal to $P_L(k)$ in linear theory, can be expressed
as integrals over $n$ powers of $P_L$ in $n^\text{th}$ order perturbation
theory (e.g.~Eq.~\eqref{eqn:pk_spt}).

In general, to give a complete statistical description of the matter
distribution at a given time, one would need to specify the entire hierarchy of
connected $n$-point correlators.  For initially Gaussian fields which are close
to linear, the higher order connection functions are small and have been
compared to simulations in \cite{Bernardeau02}.  We shall confine our attention
to the 2-point function in this paper.

The non-linear propagator (\cite{Crocce06a,Crocce06b}; also known as the
response function \cite{Valageas08}) measures the correlation between the
evolved field $\Phi_a(\vec{k};\eta)$ and the initial conditions
$\Phi_a(\vec{k}; \eta_{i})$.  It is formally defined as a functional derivative,
\begin{equation}
    \delta_D(\vec{k}-\vec{k}') G_{ab}(k;\eta,\eta_i)
        = \left\langle \frac{\delta\Phi_a(\vec{k};\eta)}{\delta\Phi_b(\vec{k}'; \eta_{i})} \right\rangle,
\end{equation}
though its significance is easier to understand from the relation
\begin{equation}
    \left\langle \Phi_a(\vec{k};\eta) \Phi_b(\vec{k}';\eta_{i}) \right\rangle
        = G_{ac}(k;\eta,\eta_i) \left\langle \Phi_c(\vec{k};\eta_i) \Phi_b(\vec{k}';\eta_i) \right\rangle ,
\end{equation}
which we shall take as a definition henceforth.
At early times or at large scales there is near-perfect correlation
($G_{ab}\approx 1$), but $G_{ab} \to 0$ on small scales as non-linear
evolution washes out the initial conditions.

Because we will make reference to it later, we also introduce here the quantity
\begin{equation} \label{eqn:nlscale}
    \Sigma^2 \equiv \frac{1}{3\pi^2} \int_0^\infty dq\,P_L(q),
\end{equation}
which characterizes the scale at which non-linearities become important.  In
the Lagrangian formalism (see below) $\Sigma^2$ gives the variance of each
component of the linear (or Zel'dovich) displacement.

\subsection{Beyond Linear Theory}

The program is now to compute the statistics of the evolved density field in terms of the 
initial density field. 
This is simple in principle but difficult
in practice, because the equations of motion are both non-linear and non-local
(in both configuration space and Fourier space).  Non-linearity forces one to
seek a perturbative solution, since exact solutions to Eqs.~\eqref{eq:fluidk}
(even if they could be found) could not be combined to construct a realistic
solution. 
A straightforward perturbative approach is hampered by
computational costs, as non-locality implies that higher order terms involve
mode-coupling integrals of ever higher dimension.

This situation has prompted a study of higher-order methods for statistical
observables like the power spectrum.  Many of these methods were borrowed from
other areas of physics (notably particle physics and fluid mechanics
\cite{L'vov95}) where
they achieved mixed success.
We review these below, highlighting the relationships between the different
methods; the methods we consider are summarized in Table~\ref{tab:kmax}.

The most straightforward approach is to define a series solution to the fluid
equations in powers of the initial density field $\delta_i$ (or equivalently,
the linearly evolved density field, $\delta_0$).
This is the basis behind {\bf standard perturbation theory\/}
(hereafter {\bf SPT};
\cite{Peebles80,Juszkiewicz81,Vishniac83,Goroff86,Makino92,Jain94});
a detailed description (including explicit expressions for $P_{ab}$ to third
order in $P_{L}$) is presented in the Appendix.

Comparisons with simulations (including those presented below) have shown that 
the domain of applicability of second order (in $P_{L}$) perturbation theory
is rather small at $z\approx 0$.
Furthermore, as we show below, going to third order is not guaranteed to
improve agreement, leading one to question the convergence properties of
such a series expansion.
If one could carry out any expansion to infinite order it would (trivially)
give the correct answer.  This however is usually not possible.  This has
led various authors to investigate ways of summing subsets of the terms
to arbitrary order in some expansion coefficent.


{\bf Renormalized perturbation theory\/} (hereafter {\bf RPT}, see
\cite{Crocce06a,Crocce06b,Crocce08}) is a variant of Dyson-Wyld resummation
(see \cite{L'vov95} for a discussion in the context of hydrodynamics)
and attempts to reorganize the perturbation expansion in terms of the
non-linear propagator and non-linear vertex to improve convergence.
In particular, if the vertex is approximated by its tree-level form then
the power spectrum can be written as an expansion in the
non-linear propagator. 
The resulting series is therefore no longer an expansion in powers of the
initial density contrast, but rather ``an expansion in orders of the
complexity of the interaction'' \cite{Wyld61}.

In \cite{Crocce06b} the dominant contributions to the non-linear propagator are
identified and summed explicitly in the high-$k$ limit, giving $G_{ab} \sim
e^{-\Sigma^2 k^2/4}$ for large $k$.  Matching this behavior with the 1-loop
propagator (valid at low $k$) gives a non-perturbative prediction for $G_{ab}$.
Substituting this propagator in the first few diagrams of the reorganized
expansion then gives a non-perturbative prediction for the power spectrum
\cite{Crocce08}.  We implemented the 1-loop and 2-loop mode-coupling
contributions as described in \cite{Crocce08}.

The above methods work at the level of the density and velocity fields; an
alternative approach is to use the fluid equations to derive equations of
motion for the power spectrum and higher order correlators directly. Such
an approach results in an infinite hierarchy of equations, which must be
somehow truncated. The {\bf closure theory\/} approach \cite{Taruya08} does so
by approximating the 3-point correlator
$\left\langle\Phi_a\Phi_b\Phi_c\right\rangle$ 
by its leading order expression in SPT.
As in \cite{Crocce06b}, $G_{ab}$ can be computed explicitly in the low-$k$ and
high-$k$ limits, and matched naturally in intermediate regimes.  The power
spectrum is then obtained order-by-order via a Born-like series expansion.

A variant of this approach (hereafter {\bf Time-RG\/} theory \cite{Pietroni08})
assumes a vanishing trispectrum to truncate the hierarchy.
The resulting equations of motion for the power spectrum $P_{ab}$ and bispectrum
$B_{abc}$ can then be numerically integrated forward in time, starting at some
sufficiently early redshift $z_i$ (where $P = P_L$ and $B = 0$).
Since the time evolution is performed numerically, the method also allows the
proper treatment of models where the linear growth factor is scale-dependent
(e.g.\ models with quintessence or massive neutrinos \cite{Lesgourgues09}).
This approach may be seen as a generalization of the
{\bf renormalization group perturbation theory} (hereafter {\bf RGPT}) of
\cite{McDonald07}, which is an attempt to regulate the relative divergence
of 1-loop SPT using renormalization group methods.

In \cite{Valageas01,Valageas02,Valageas04} a path-integral
formulation of the Vlasov equation is developed in terms of the distribution
function $f(\vec{x},\vec{p},t)$.  In \cite{Valageas07} a similar technique is
applied to the fluid equations (Eq.~\eqref{eq:fluidphi}).  The key insight here is
that statistical observables like the power spectrum may be obtained by taking
functional derivatives of an appropriately constructed path integral (the
generating functional).  Straightforward perturbative evaluation of the
generating functional reproduces the results of SPT, whereas applying large-$N$
expansion techniques and truncating at fixed order in $1/N$ leads to
approximations for the power spectrum and propagator.  These approximate
solutions agree with SPT up to a fixed order in $P_0$, but also include
non-perturbative contributions corresponding to infinite partial resummations
of the standard expansion.  We focus attention on the steepest-descent method
of \cite{Valageas07} (hereafter {\bf Large-N}), 
as it is considerably easier to implement than the 2PI
effective action method.

{\bf Lagrangian resummation theory} \cite{Matsubara08,Matsubara08b} is an
extension of the well-developed Lagrangian perturbation theory.  Lagrangian
perturbation theory (hereafter LPT; see \cite{Buchert92,Buchert93,Buchert94})
has received less attention recently than its Eulerian counterpart as a method
for investigating non-linear structure growth, partly because the Lagrangian
picture breaks down once shell-crossing occurs.  However, recent work
\cite{Matsubara08} has demonstrated that Lagrangian perturbation theory not
only reproduces the SPT power spectrum at the lowest non-trivial order, but
with a slight modification also yields a non-perturbative prediction for the
power spectrum that corresponds to resumming an infinite set of terms in the
standard expansion.  We review LPT and the cumulant expansion in Appendix
\ref{sec:lptreview}.

\section{Simulations} \label{sec:sims}

In order to assess how well the perturbative expansions are doing, we need a
reference for any given cosmology.  We use a new set of large dynamic range
N-body simulations well suited to this purpose.  These computer programs
simulate the same basic physical system (a collisionless matter `fluid'
interacting only through gravity) that the perturbative methods attempt to
describe; hence the results of the two methods, though arrived at very
differently, are directly comparable.

We have elected to investigate several different cosmologies, in an attempt to
better identify where and why various analytic techniques succeed and/or fail.
For simplicity we consider only flat models in the CDM family.
We will highlight two: the first in which a cosmological constant dominates
the late-time evolution and which is close to the best-fit cosmology
($\Lambda$CDM: $\Omega_M=0.25$, $\Omega_b h^2 = 0.0224$, $h=0.72$, $n=0.97$ and $\sigma_8=0.8$)
and an extreme model
($c$CDM: $\Omega_M=1$, $\Omega_b h^2 = 0.1$, $h=0.5$, $n=1$, $\sigma_8=1$)
with a critical density in matter and a larger present-day normalization
which emphasizes the effects of non-linearity and the erasure of baryon
acoustic oscillations through mode coupling.

For each cosmology the transfer function, $T(k)$, was computed by evolving the
coupled Boltzmann, fluid, and Einstein equations using the publicly available
package CAMB ({\tt http://www.camb.info}).  The resulting power spectra were
then used both as input to the perturbative methods and to generate initial
conditions for the N-body simulations (Table \ref{tab:delta_lin} gives the
amplitude of the dimensionless power at some fiducial wavenumbers).

\begin{table}
\begin{tabular}{c|cccc} \hline
$k$ & \multicolumn{2}{c}{$\Delta_L^2(z=1)$}
    & \multicolumn{2}{c}{$\Delta_L^2(z=0)$}  \\
    & $\Lambda$CDM & $c$CDM & $\Lambda$CDM & $c$CDM \\ \hline
0.05 & 0.03 & 0.03 & 0.09 & 0.14 \\
0.10 & 0.11 & 0.09 & 0.27 & 0.36 \\
0.15 & 0.19 & 0.22 & 0.49 & 0.90 \\
0.20 & 0.29 & 0.27 & 0.72 & 1.07 \\
0.25 & 0.37 & 0.38 & 0.94 & 1.51 \\
\hline
\end{tabular}
\caption{\label{tab:delta_lin}The value of the dimensionless, linear power
spectrum at $z=1$ and $z=0$ at several fiducial wavenumbers for our two
example cosmologies. $k$ is given in $h\,\text{Mpc}^{-1}$.}
\end{table}

A number of numerical issues need to be addressed in order to ensure that our
simulations provide an adequate reference.  Our workhorse simulations each
employ $1024^3$ equal mass dark matter particles in a periodic, cubical box
of side length $2\,h^{-1}$Gpc.  By employing such large volumes we are highly
insensitive to the periodicity of the box, which represents a fair sample of
the Universe \cite{CoyoteI}.
There is very little power at the fundamental mode, even at $z=0$:
$\Delta^2(k_f,z=0)<10^{-4}$.  The lowest few modes obey linear growth to
sub-percent accuracy and we run enough different realizations to ensure that
the spectrum at the scales of interest is well determined.
The large number of particles ensures that the spectrum is well converged
for the $k$-modes of interest, which we checked explicitly by comparing
simulations of different box sizes.
The simulations are evolved from $z_i = 100$, with the particles
perturbed from an initially uniform grid using the Zel'dovich approximation.
The rms particle move was about 5\% of the mean interparticle spacing.
Comparison with second order Lagrangian perturbation theory initial conditions
showed that this starting redshift is sufficently high that transients from
the Zel'dovich start are irrelevant for the scales and redshifts of interest.

Most of the evolutions were performed with a parallel particle-mesh code.  To
cross-check our results we used two high force resolution N-body codes: the
TreePM code \cite{White02} and Gadget-II \cite{Gadget2}.
These have each been tested against a suite of other codes
\cite{Evrard08,Heitmann08,CoyoteI}, with very good agreement.
We ran a subset of our simulations using all three codes to quantify the
level of precision for the box size and particle loading of relevance here.
With its default time stepping, the TreePM code produces dark matter power
spectra in agreement with those from Gadget-II to better than $0.2\%$ out
to $k\simeq 1\,h\,{\rm Mpc}^{-1}$ and to $\mathcal{O}(10^{-4})$ for
$k<0.1\,h\,{\rm Mpc}^{-1}$.  However these runs prove to be quite time
consuming.  If we set the time step in the TreePM code to
\begin{equation}
  \left( \delta \ln a\right)^{-2} = \left[\frac{1}{0.05}\right]^2
  + \left[\frac{a}{0.01}\right]^2,
\end{equation}
which evolves from 5\% steps at early times to 1\% steps as $a \to 1$, we find
a shortfall of power of approximately $1\%$ at
$k\simeq 1\,h\,{\rm Mpc}^{-1}$ but very little difference for
$k<0.1\,h\,{\rm Mpc}^{-1}$.
We choose the same time stepping for the particle-mesh code, which results
in very short run times allowing an ensemble of simulations to be performed.
With this step the particle-mesh power spectra show a significant deficit
of power (compared to TreePM or Gadget-II) beyond
$k\approx 0.7\,h\,{\rm Mpc}^{-1}$ but for $k<0.2\,h\,{\rm Mpc}^{-1}$,
the region of interest here, the agreement is better than $1\%$.

To compute the power spectrum at different output times the particles were
binned onto a regular, Cartesian grid using charge-in-cell assignment
\cite{Hockney88}
and the resulting density field was Fourier transformed.  The Fourier modes
were squared, corrected for the gridding by dividing by the Fourier transform
of the charge assignment scheme, and binned into bins equally spaced in
$\log k$.  The average of $\Delta^2(k)$ was assigned to the average $k$ in
the bin and shot-noise was subtracted assuming it was Poisson.
The binning introduces artifacts at low $k$, where the sampling on the
grid is sparse and the dimensionless power spectrum is steep, but these are
small for the scales of most relevance to us.
Similarly there is some evidence that the shot-noise in simulations is not
scale-invariant (Poisson), but the correction is negligibly small on the
scales of interest here.

The non-linear propagator was computed by cross-correlating the initial density
field with the final one \cite{Crocce06b}.  Similar to the power spectrum, this
quantity is obtained by Fourier transforming both fields, multiplying their
Fourier coefficients, correcting for gridding, and then binning the results.

The velocity statistics are more problematic, because while the density and
momentum fields must vanish where there are no tracer particles, the same is
not true of the velocities.  Thus estimates of the velocity field must employ a
smoothing technique.  Similarly the velocity field is more sensitive to finite
force resolution.  On the other hand comparison of the velocity fields with the
density fields is less sensitive to finite volume scatter.  For this reason we
use a different set of simulations, with more particles (up to 3 billion) in
smaller boxes ($1.25\,h^{-1}$Gpc down to $720\,h^{-1}$Mpc) evolved with the
TreePM code, for the velocity statistics.  Comparison with different smoothing
schemes, box sizes and particle loadings show that with these choices our
results are well converged on the scales of interest
\cite{VelFisher}.

\section{Comparison} \label{sec:compare}

\subsection{The Power Spectrum}

\begin{figure*}
\resizebox{3.5in}{!}{\includegraphics{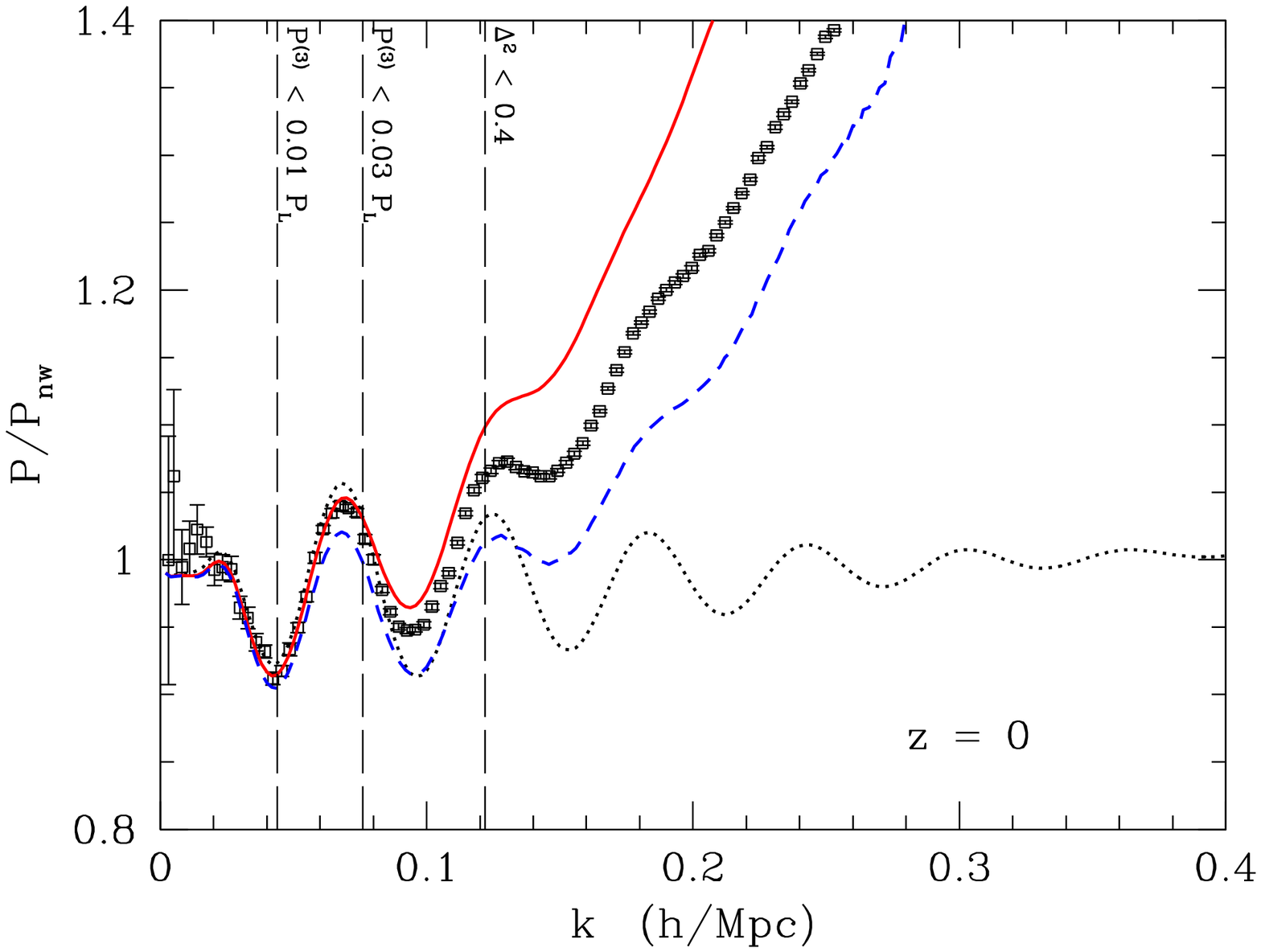}}
\resizebox{3.5in}{!}{\includegraphics{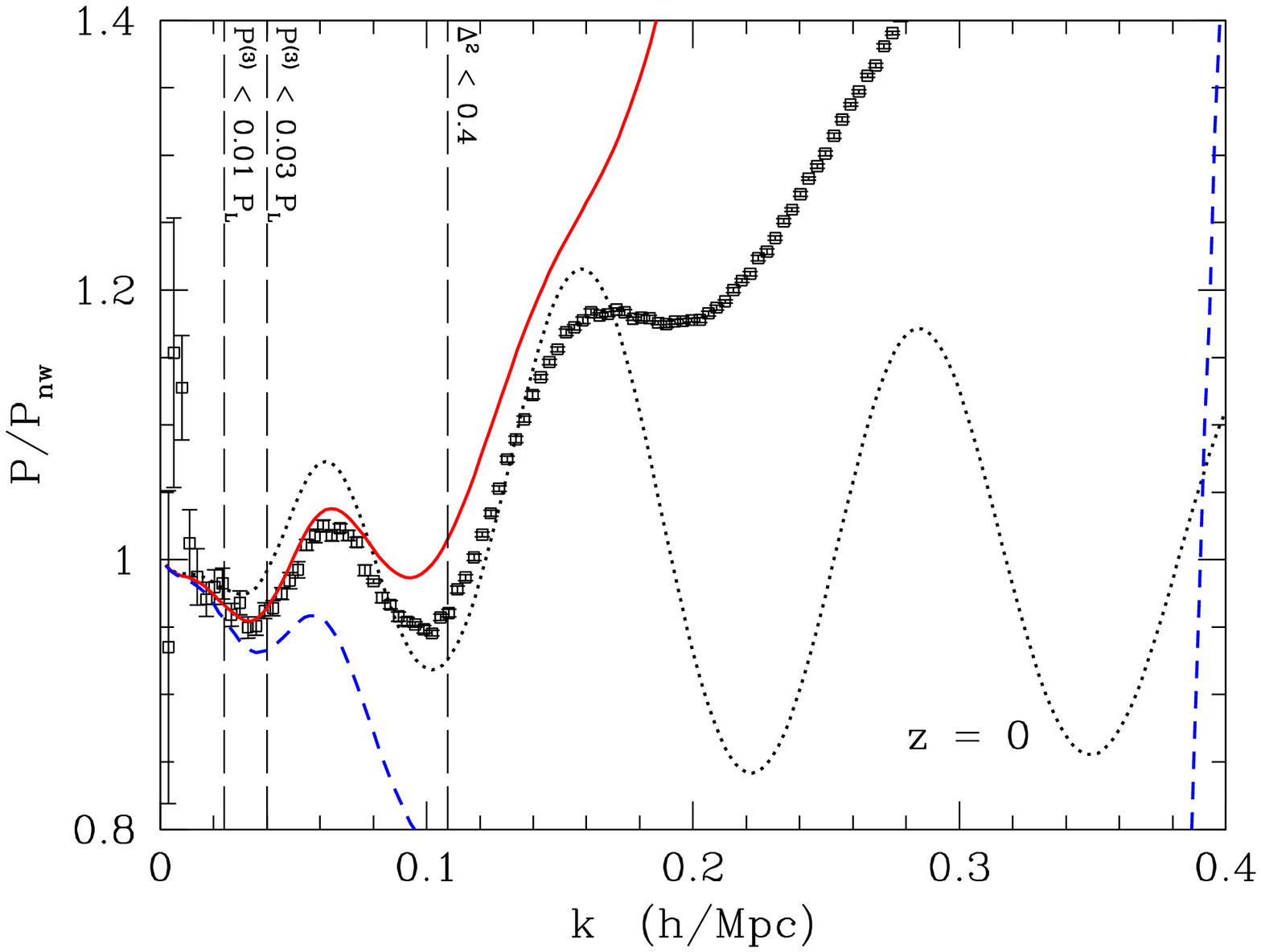}}
\resizebox{3.5in}{!}{\includegraphics{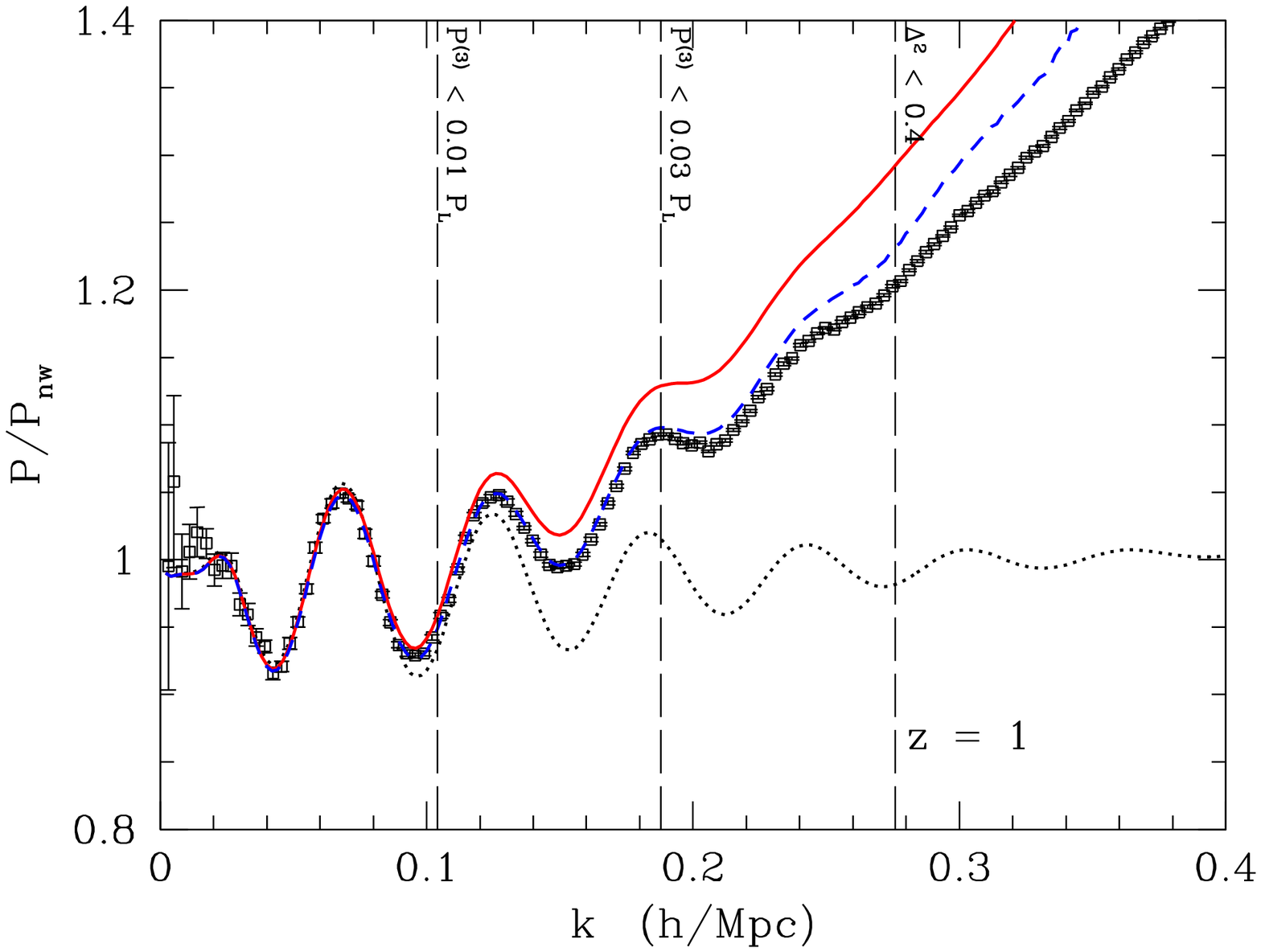}}
\resizebox{3.5in}{!}{\includegraphics{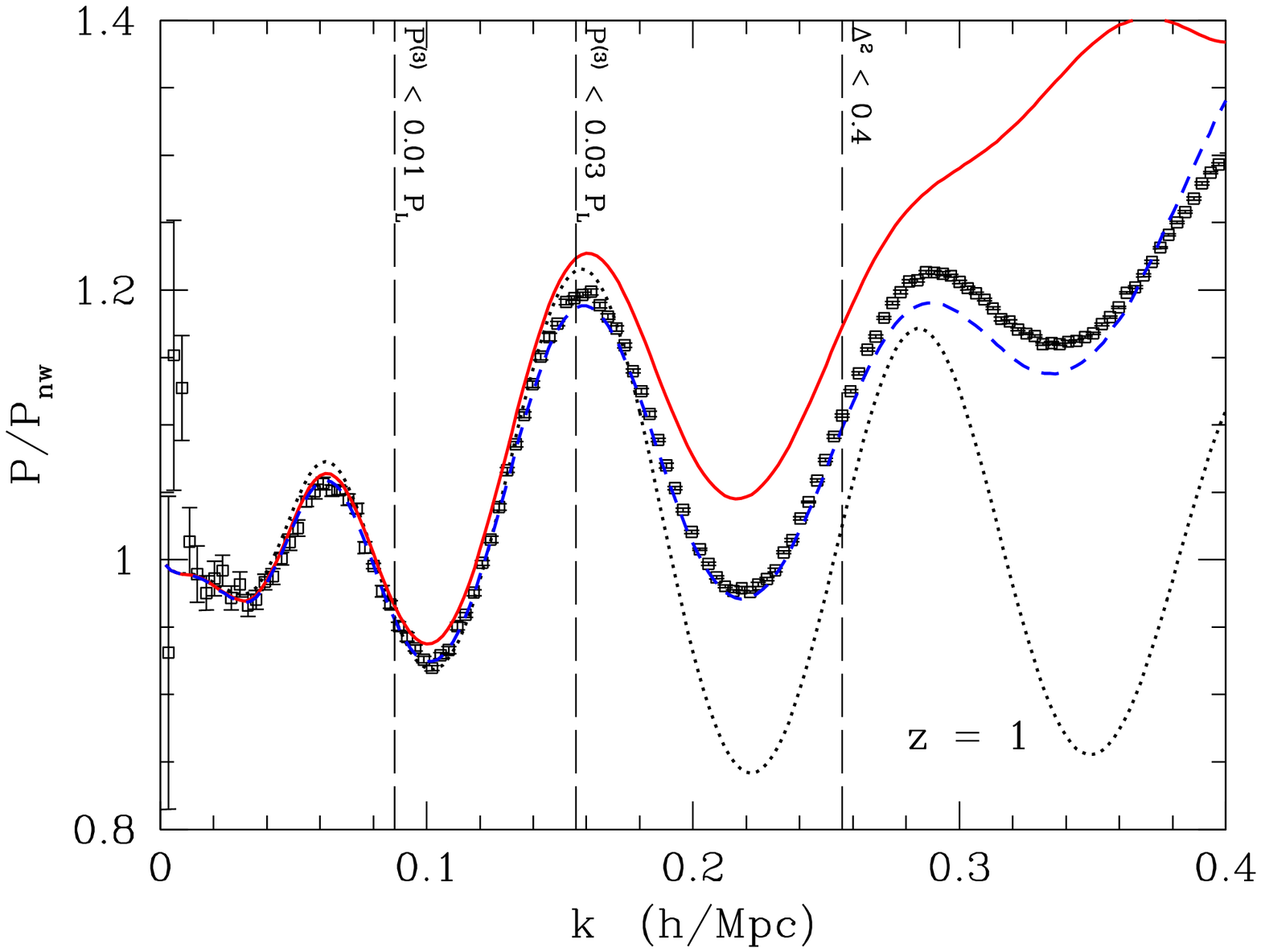}}
\caption{SPT power spectrum at linear (black; dotted), 1-loop (red; solid),
and 2-loop (blue; dashed) order.  The squares with error bars show the mean
and error from our N-body simulations.  The four panels show $\Lambda$CDM
(left) and $c$CDM (right) at redshifts 1 (top) and 0 (bottom).
Each curve has been divided by the no-wiggle power spectrum of
\protect\cite{Eisenstein98} to reduce the dynamic range.  We also indicate
the domain of validity of 1-loop SPT according to the heuristic prescription
of \protect\cite{Jeong06} ($\Delta^2<0.4$), and according to the criterion
$P^{(3)} < \alpha\, P_L$ for $\alpha = 0.01, 0.03$.}
\label{fig:spt}
\end{figure*}

We begin our analysis by comparing the predictions of SPT against our
simulation results.  Figure \ref{fig:spt} shows the linear theory, 1-loop SPT,
and 2-loop SPT power spectrum for $\Lambda$CDM and $c$CDM.  While 2-loop SPT is
a marked improvement over 1-loop SPT at $z = 1$, it's actually worse than
1-loop at $z = 0$.  The effect is most apparent in $c$CDM, which has larger
$\sigma_8$ and $\Omega_b$.  This break-down in standard perturbation theory is
not entirely surprising: since the $n^\text{th}$ order term in SPT goes like
$D^{2n}(z)$, at any given scale one expects higher order terms to become
comparable in magnitude to lower order terms at sufficiently late times.  Our
results suggest that at BAO scales (roughly $k = 0.05-0.25\,h\,\text{Mpc}^{-1}$)
the break-down occurs between $z = 1$ and $z = 0$.

A common heuristic prescription dictates that 1-loop SPT can be trusted to 1\%
for wavenumbers satisfying $\Delta_L^2(k) \lesssim 0.4$ \cite{Jeong06}.  On the
other hand a strict application of perturbation theory implies that 1-loop SPT
can be trusted to 1\% for wavenumbers where the 2-loop contribution is 1\% of
linear theory.  In Figure \ref{fig:spt} we indicate the predicted domain of
validity of 1-loop SPT according to these two criteria.  For comparison we also
indicate where the 2-loop contribution is within 3\% of linear theory.  One
sees that the agreement with simulations is slightly better than what our more
rigorous criterion suggests.
For instance for $\Lambda$CDM at $z=0$, $\Delta_L^2 = 0.4$ at
$k_* \approx 0.12\,h\,\text{Mpc}^{-1}$.  At this wavenumber 1-loop SPT
overshoots the reference spectrum by about 3\%, whereas 2-loop SPT undershoots
the reference spectrum by 5\%.  For $c$CDM at $z=0$ the situation is much worse,
with 1-loop SPT overshooting by only 6\% at $k_* \approx 0.11\,h\,\text{Mpc}^{-1}$,
but 2-loop SPT undershooting by almost 20\%.

\begin{figure*}
\resizebox{3.5in}{!}{\includegraphics{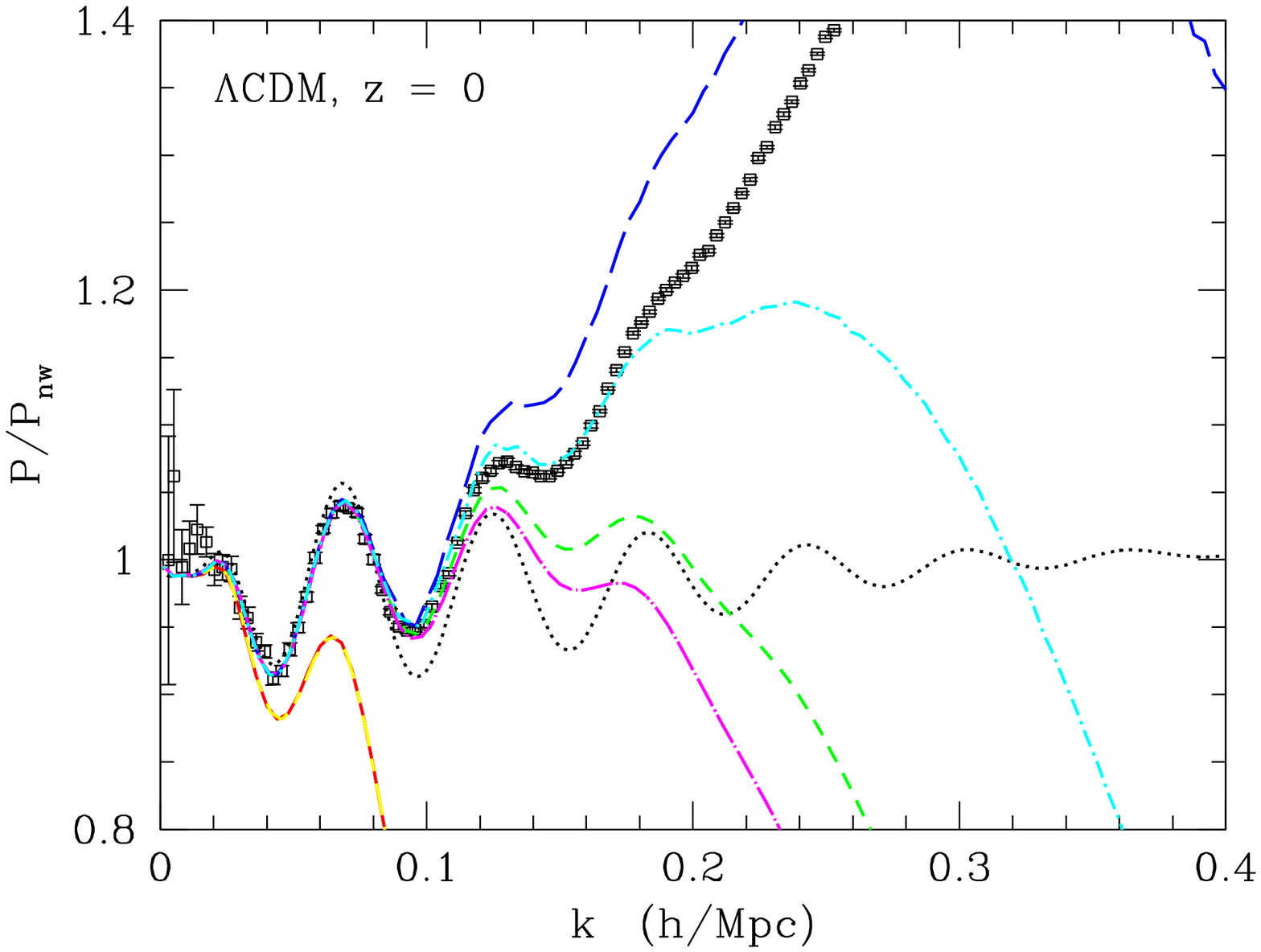}}
\resizebox{3.5in}{!}{\includegraphics{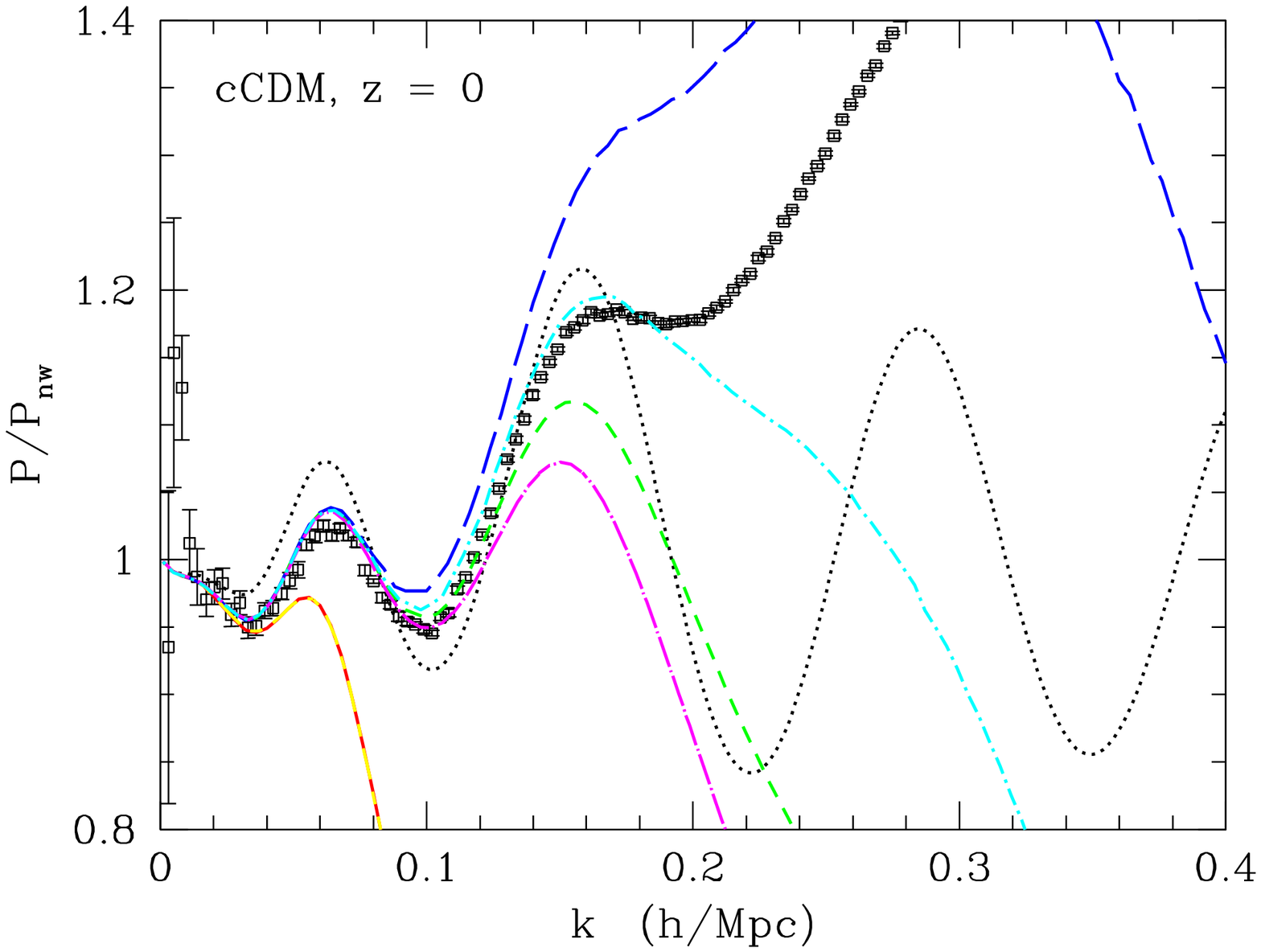}}
\resizebox{3.5in}{!}{\includegraphics{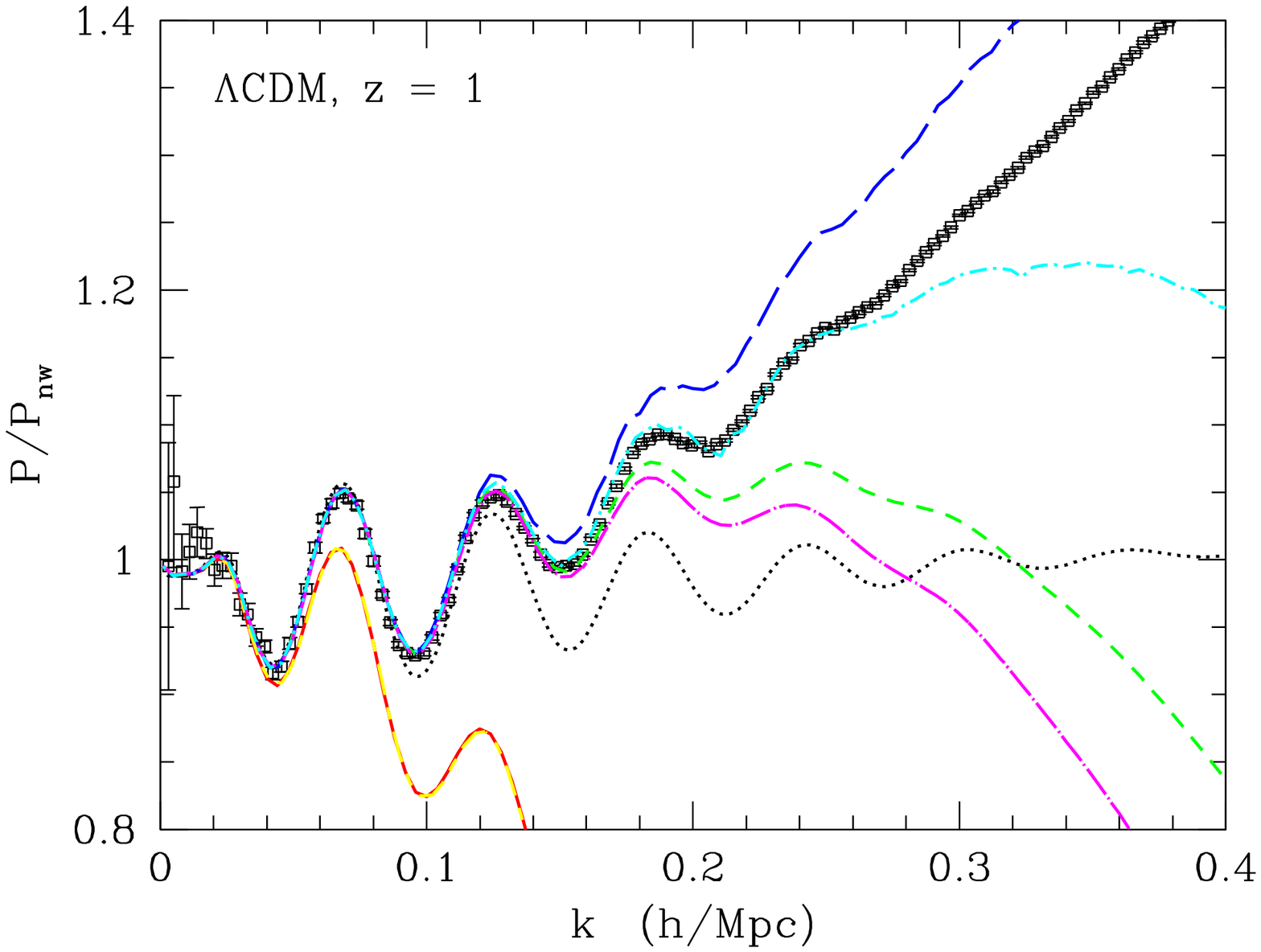}}
\resizebox{3.5in}{!}{\includegraphics{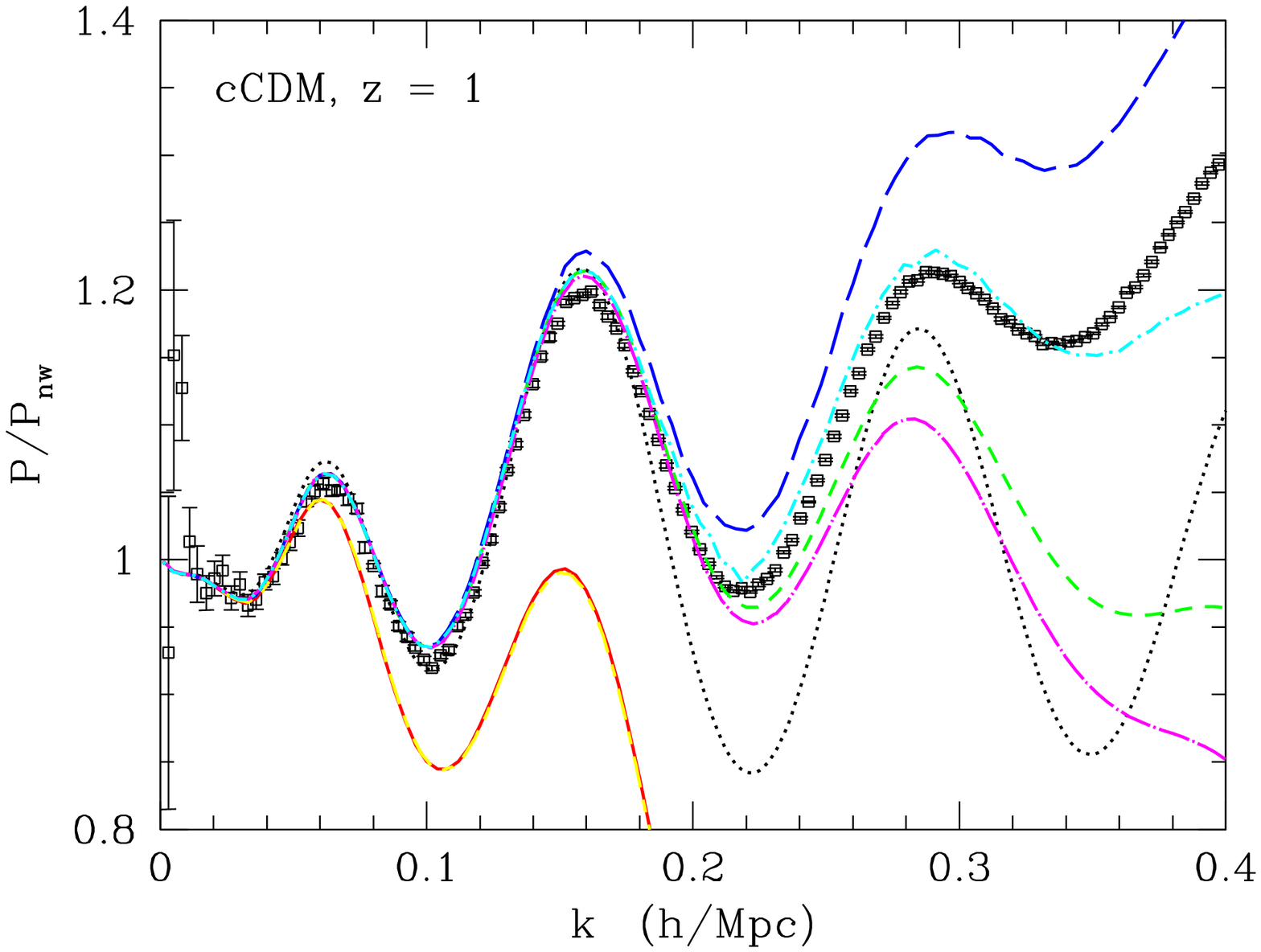}}
\caption{Comparison of the tree-level, 1-loop and 2-loop power spectrum from
RPT and closure theory, for $\Lambda$CDM (left) and $c$CDM (right).
Each curve has been divided by the no-wiggle power spectrum of
\protect\cite{Eisenstein98} to reduce the dynamic range.
The (black) dotted line is linear theory, the (red) solid line is tree-level
RPT, the (green) dashed line is 1-loop RPT, the (blue) long-dashed line is
2-loop RPT, the thick (yellow) short-long dashed line is tree-level closure,
the (magenta) dot-long dashed line is 1-loop closure, and the (cyan) dot-dashed
line is 2-loop closure.}
\label{fig:2loop}
\end{figure*}

RPT and closure theory have also been developed to two loops.  Given the above
conclusions about SPT, it is natural to make the same comparison between the
1-loop and 2-loop predictions from RPT and closure theory.  In Figure
\ref{fig:2loop} we show the matter power spectrum for these theories at tree,
1-loop, and 2-loop order for both $\Lambda$CDM and $c$CDM.
For closure theory it appears that going to 2-loop order extends the range
of agreement with simulations, although the wiggles of the power spectrum are
not matched in detail. 
For RPT, as with SPT, the 2-loop result is systematically high, whereas the
1-loop result performs fairly well below $k \approx 0.15\,h\,\text{Mpc}^{-1}$.
Agreement with simulations can be improved
by changing the damping scale in the propagator.  In \cite{Crocce08} the
damping scale was modified by calculating $\Sigma$ with the linear theory
expression (Eq.~\ref{eqn:nlscale}), but using the non-linear power
spectrum and integrating only up to $k=4\,k_{\rm nl}$.
This leads to a $\sim 10\%$ additional suppression of $G(k)$ and hence $P(k)$
on the relevant scales, bringing the theory into better agreement with
simulations \cite{Crocce08}.  At present this correction has not been
derived from first principles and we have not included it, but it appears
that improvements in this direction could be important.

\begin{figure*}
\resizebox{3.5in}{!}{\includegraphics{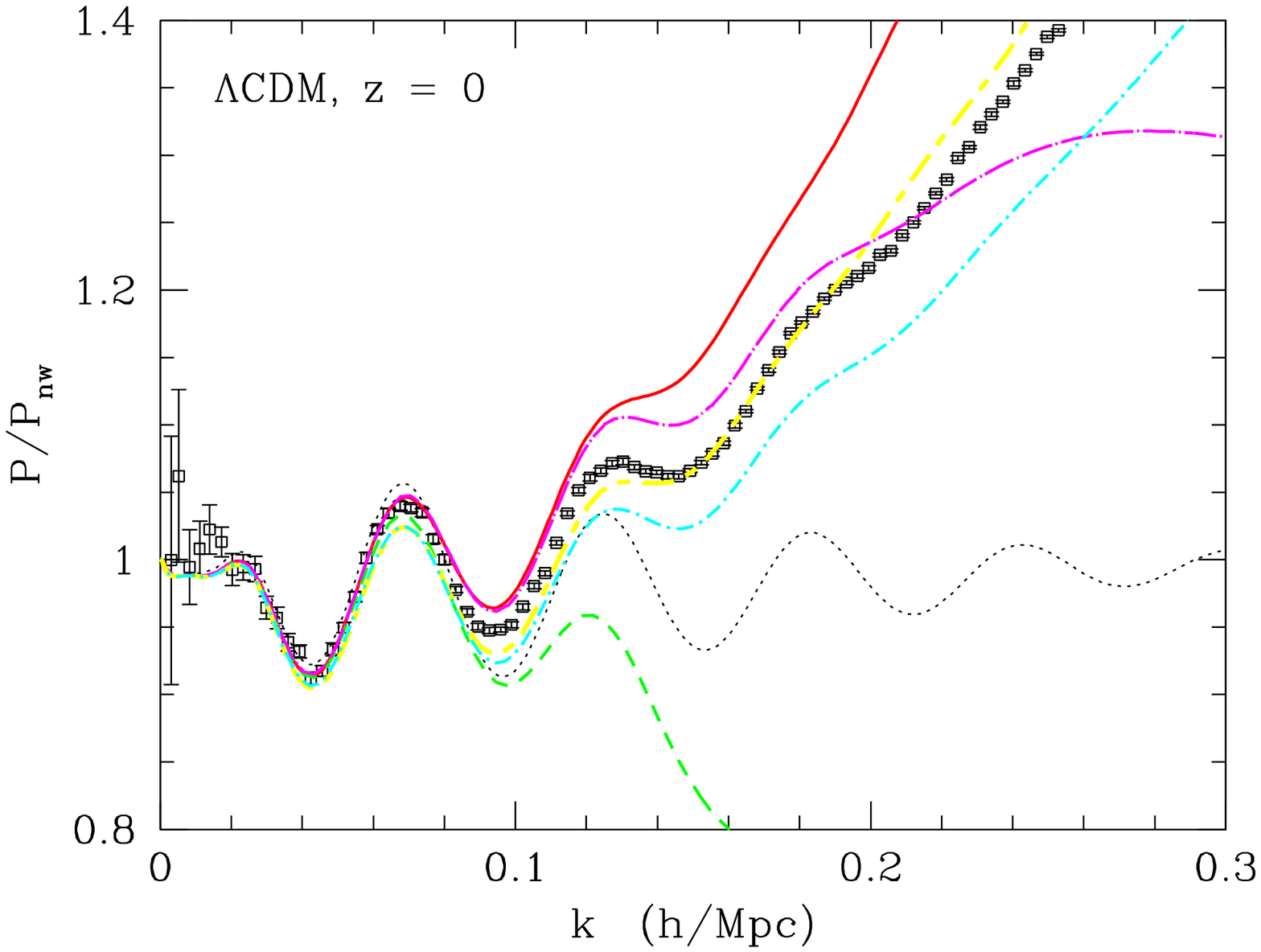}}
\resizebox{3.5in}{!}{\includegraphics{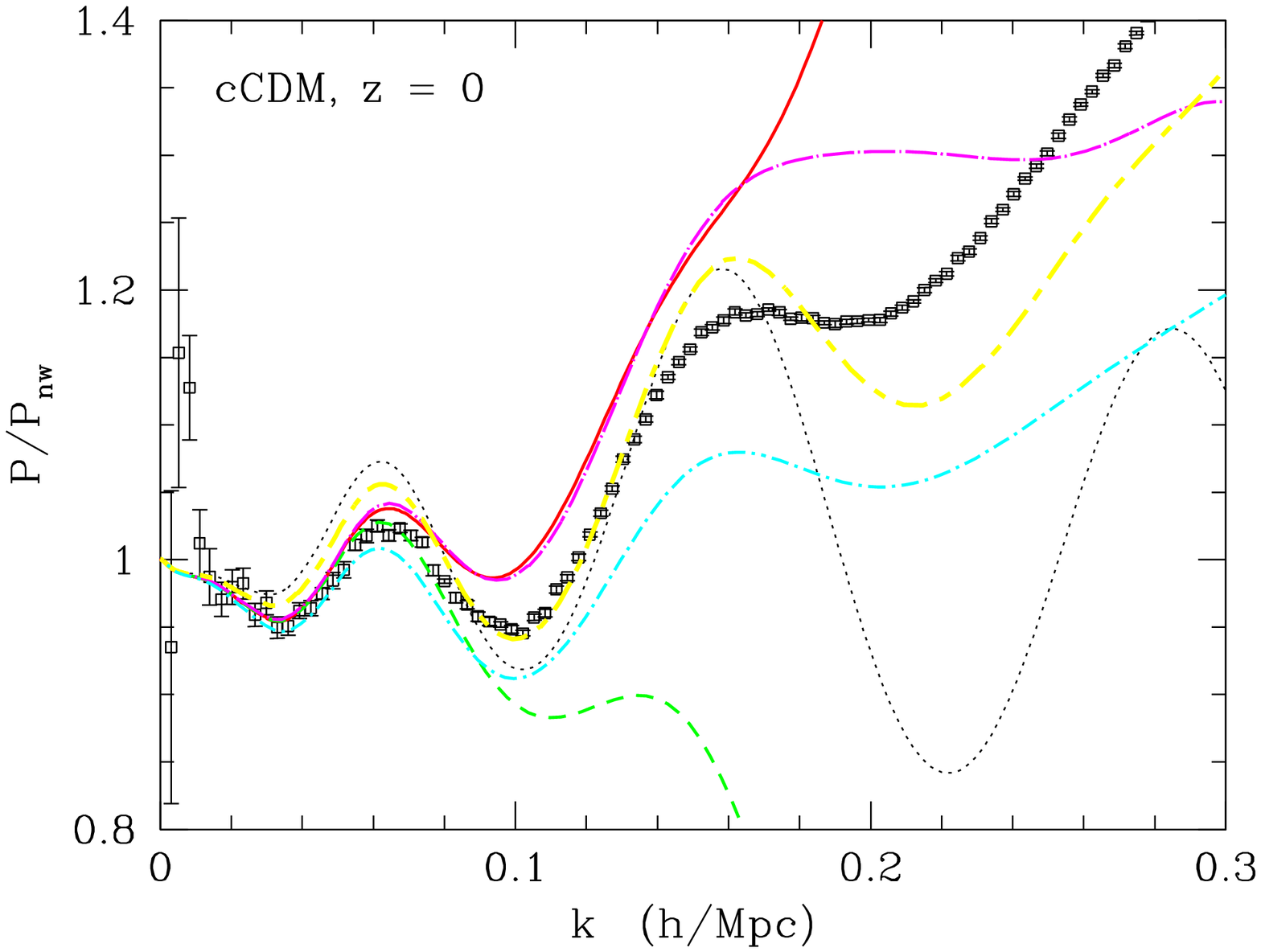}}
\resizebox{3.5in}{!}{\includegraphics{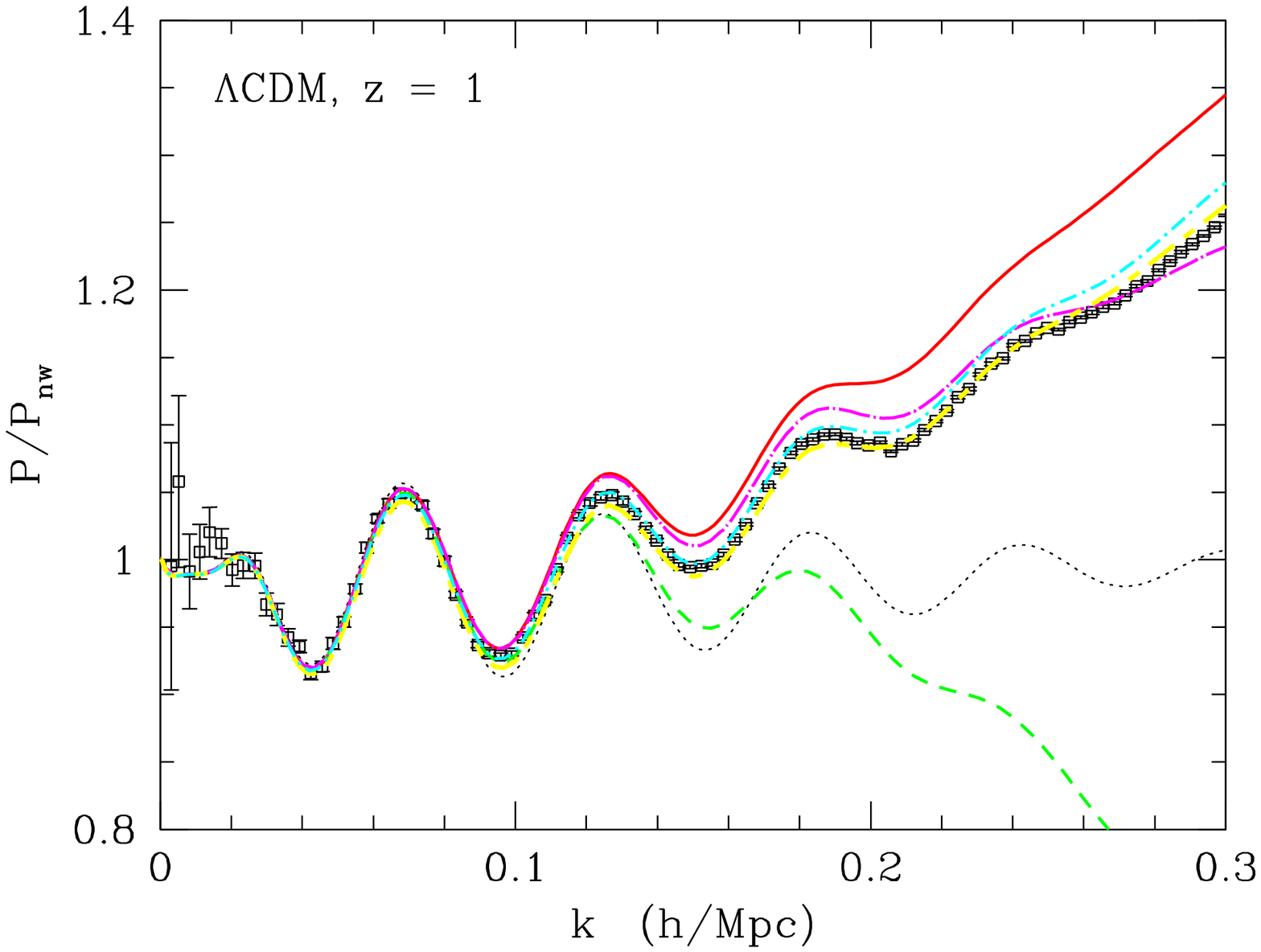}}
\resizebox{3.5in}{!}{\includegraphics{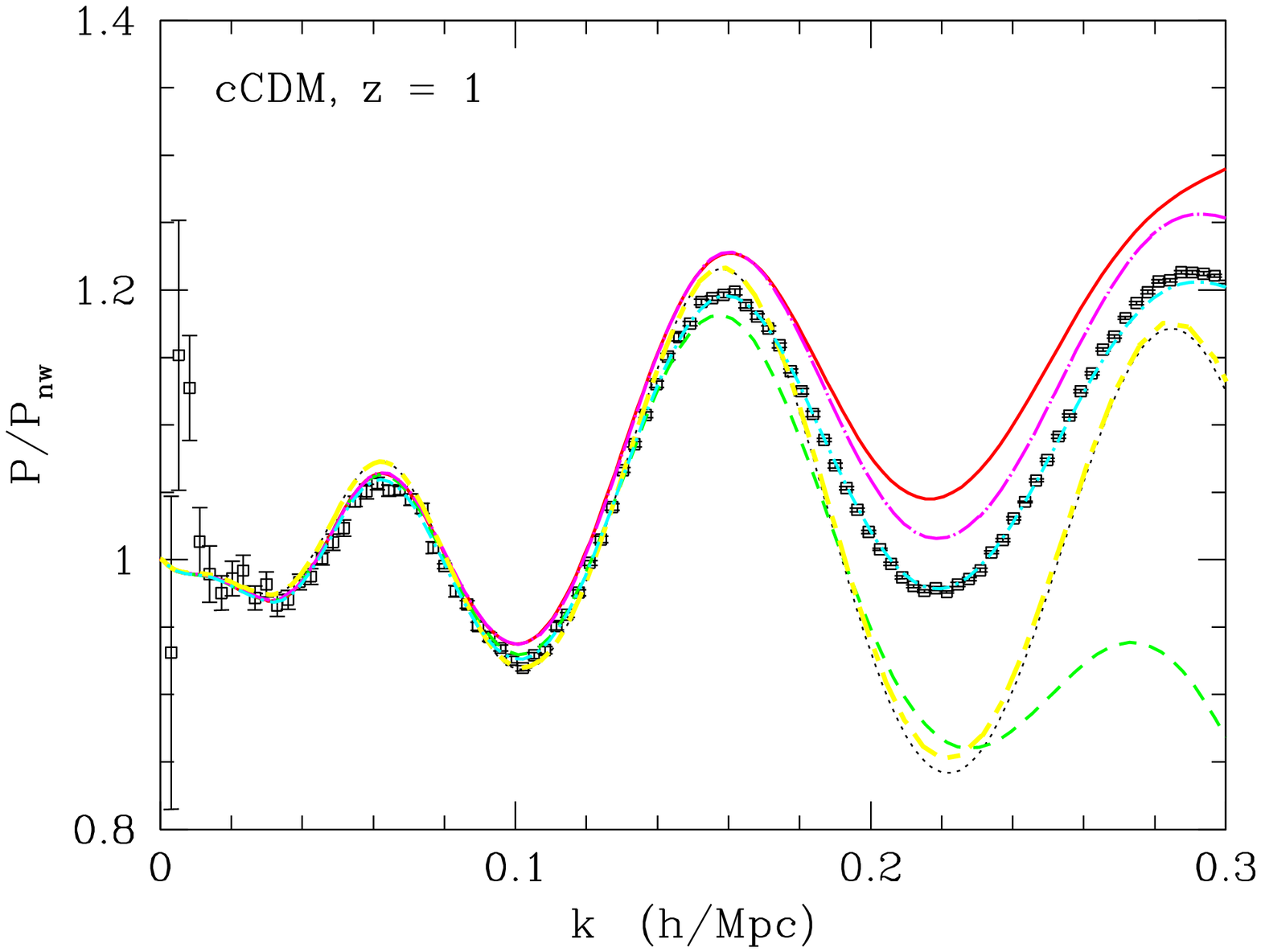}}
\caption{Comparison of the power spectrum for the remaining theories.  Each
curve has been divided by the no-wiggle power spectrum of
\protect\cite{Eisenstein98} to reduce the dynamic range.
The (red) solid line is 1-loop SPT, the (magenta) dot-long dashed line is
large-$N$ theory, the (green) dashed line is Lagrangian resummation, the
thick (yellow) short-long dashed line is time-RG theory, and the (cyan)
dot-dashed line is RGPT.}
\label{fig:others}
\end{figure*}

Figure \ref{fig:others} shows the predicted power spectrum for the remainder of
the theories that we consider in this work.  With Figures \ref{fig:spt} and
\ref{fig:2loop}, these figures give an overview of the agreement between
our N-body simulations and the perturbation theories for $\Lambda$CDM and
$c$CDM.  Some of the trends can be seen easily in these figures, and are generic
across cosmologies and redshifts.  For instance 1-loop SPT, which is the same
as 1-loop LPT, always overpredicts $P(k)$ at high $k$.  Lagrangian resummation
theory on the other hand is much too strongly damped beyond the first wiggle.
Large-$N$ theory more or less traces 1-loop SPT before turning over, while
time-RG theory and RGPT follow the general trends of the N-body data without
fitting any particular feature precisely.  (Note that the nearly perfect
agreement between RGPT and simulations for $c$CDM at $z=1$ is likely spurious,
as this level of agreement is not seen for other cosmologies or at other
redshifts.)
RPT and closure give nearly identical tree-level predictions, and very similar
1-loop predictions for $P(k)$.  Closure theory appears to benefit greatly from
going to 2-loop order, whereas for RPT even at $z=1$ it appears that 2-loop does
worse than 1-loop.

While we have run many realizations of each cosmology to reduce run-to-run
variance, one sees in Figures \ref{fig:spt}, \ref{fig:2loop} and
\ref{fig:others} that the N-body data are still noisy at low $k$, which makes
it difficult to make quantitative statements about the performance of the
perturbation theories.  To overcome this we define a
`reference spectrum' which interpolates the N-body results at high and
intermediate $k$ with the 1-loop SPT calculation at low $k$.
This eliminates the large scatter from the finite number of modes in the
simulations and any biases from the finite bin sizes at low $k$, while still
retaining the information from the simulations at larger $k$.  This gives a
smooth function, defined for all $k$, which can be used as a reference
to make a quantitative comparison.
Given the large number of simulations we have run, the uncertainty in the
N-body results is small before perturbation theory becomes invalid and we
can see a significant range of $k$ for which theory and simulation agree well.
This makes our final results insensitive to how the matching is done.
Our recipe for producing a reference spectrum is to treat both the N-body
results and 1-loop SPT as independent measurements of the true power spectrum,
with errors given by the run-to-run variance within a wavenumber
bin \footnote{We bin the model into the same finite-width bins as the N-body
data when making the comparison to the latter.} in the former case,
and by the 2-loop SPT term in the latter case.
Then the reference spectrum at any given $k$ is defined by fitting a
polynomial to all available measurements within a small wavenumber range
$[k-\Delta k,k+\Delta k]$ and evaluating that polynomial at $k$.
For simplicity we chose to fit to a cubic with
$\Delta k=0.01\,h\,{\rm Mpc}^{-1}$,
though the resulting reference spectrum is rather insensitive to these
choices.


\begin{figure}
\resizebox{3.5in}{!}{\includegraphics{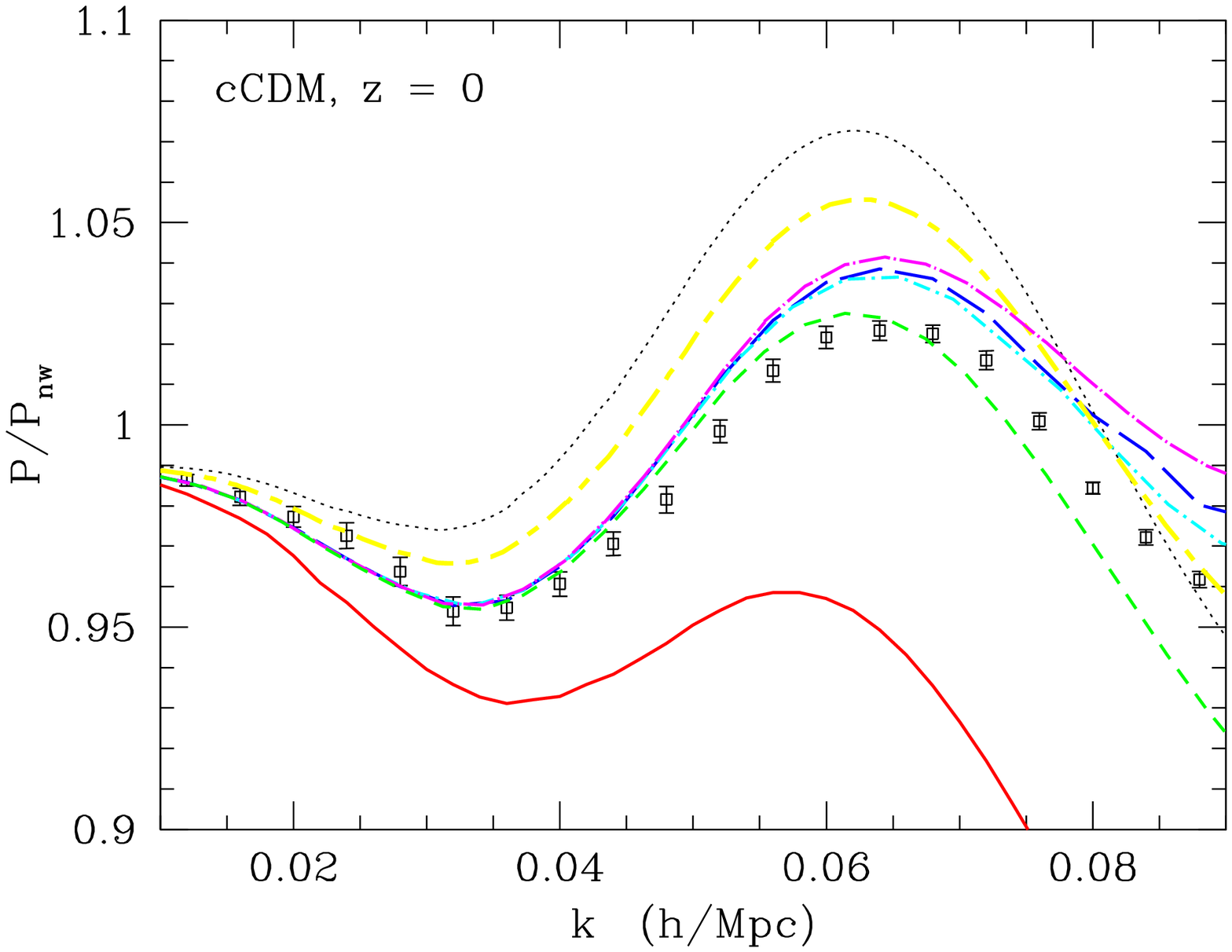}}
\caption{Comparison between analytic models for $P(k;z=0)$ and the reference
spectrum (Section \ref{sec:compare}) for model $c$CDM, focusing on large
scales.  Each curve has been divided by the no-wiggle power spectrum of
\protect\cite{Eisenstein98} to reduce the dynamic range.  The points with error
bars are the `reference spectrum' defined in the text.  The (black) dotted line
is linear theory, the (red) solid line is 2-loop SPT, the (blue) long-dashed
line is 2-loop RPT, the (green) short-dashed line is Lagrangian resummation,
the (cyan) dot-dashed line is 2-loop closure theory, the thick (magenta)
dot-long dashed line is the large-N expansion, and the thick (yellow)
short-long dashed line is time-RG theory.}
\label{fig:previr}
\end{figure}

All of the theories beyond linear correctly predict the `dip' below linear
theory which can be most clearly seen in Figure \ref{fig:previr} around
$k\simeq 0.04\,h\,\text{Mpc}^{-1}$.  This is sometimes referred to as
pre-virialization, and arises because the non-linear growth of the density and
velocity fields is slower than linear on scales where the effective spectral
index is more positive than (about) $-1.5$ (see \cite{Bernardeau02} for further
discussion).  In this region use of any of the methods provide significant
improvements over linear theory.

\begin{figure*}
\resizebox{3.5in}{!}{\includegraphics{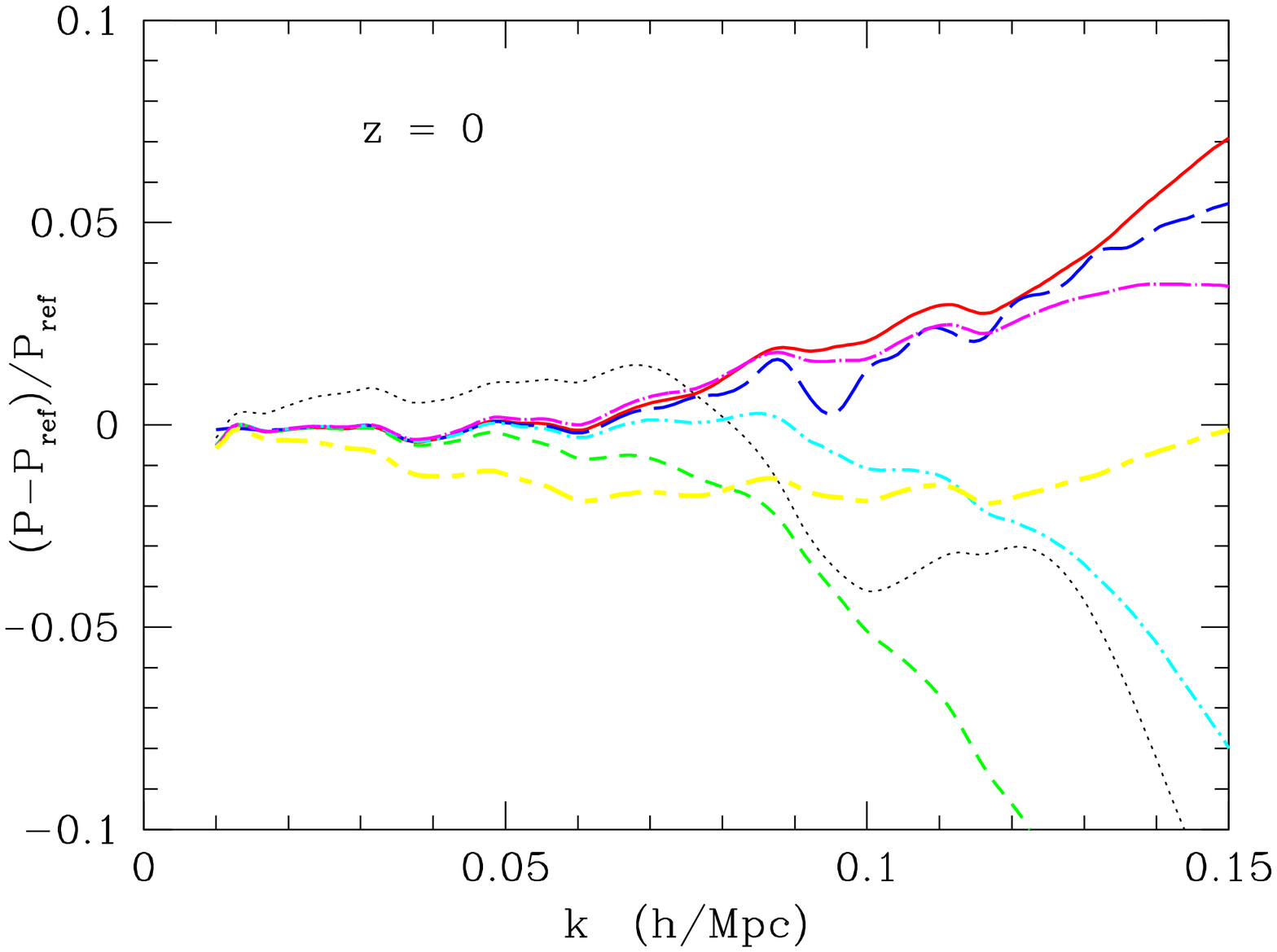}}
\resizebox{3.5in}{!}{\includegraphics{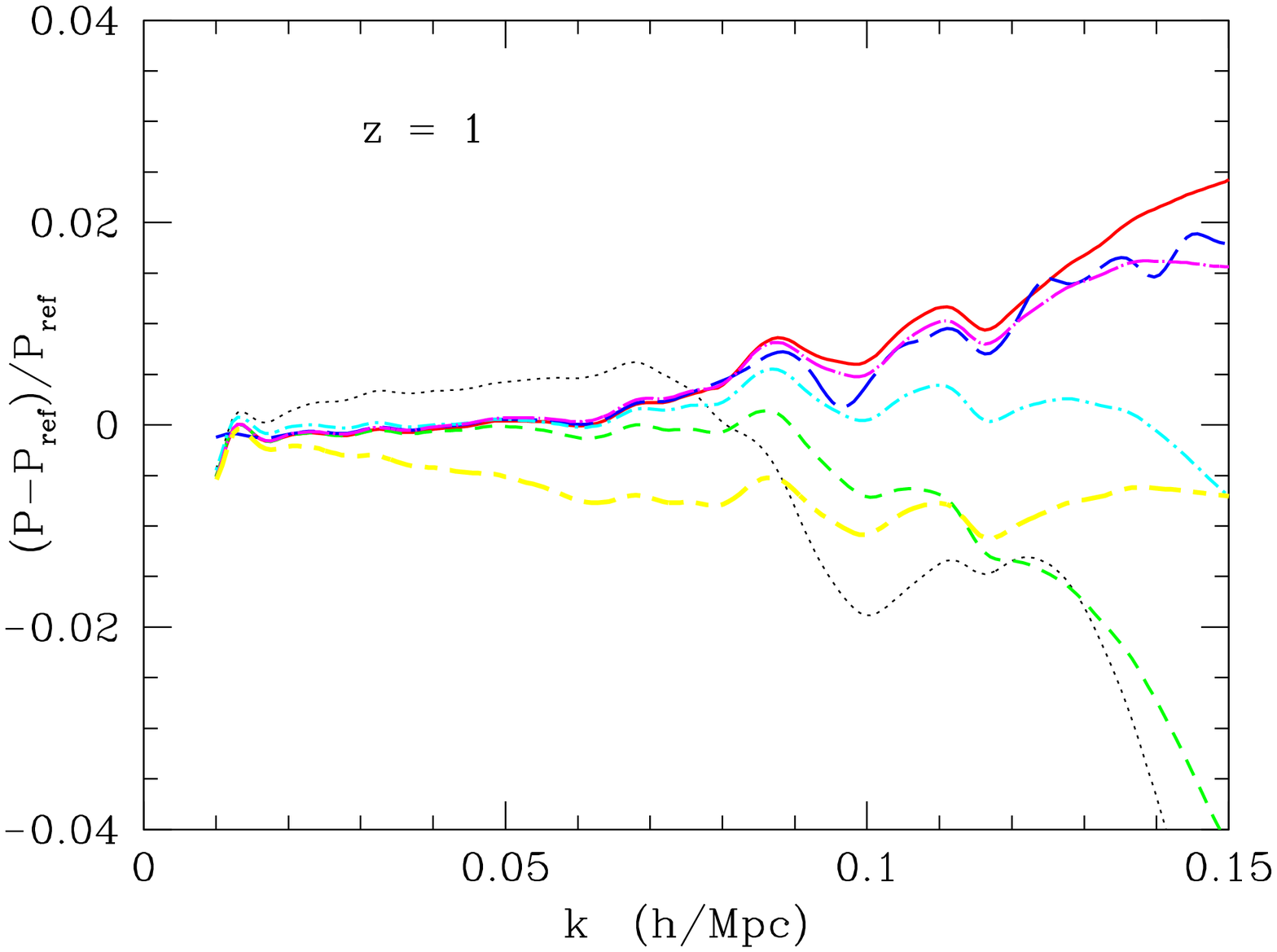}}
\caption{The fractional deviation of each method from the reference spectrum,
for $\Lambda$CDM at $z=0$ (left) and $z=1$ (right).  This figure focuses on the region
$k<0.15\,h\,\text{Mpc}^{-1}$ where linear theory is inadequate but higher
order methods are still viable.  As in Figure \protect\ref{fig:previr} the (black)
dotted line is linear theory, the (red) solid line is 2-loop SPT, the (blue)
long-dashed line is 2-loop RPT, the (green) short-dashed line is Lagrangian
resummation, the thick (cyan) dot-short dashed line is 2-loop closure theory the thick
(magenta) dot-long dashed line is the large-$N$ expansion, and the thick
(yellow) short-long dashed line is time-RG theory.}
\label{fig:pk_diff}
\end{figure*}

\begin{table}
\begin{tabular}{lc|ccccc} \hline
       &   & \multicolumn{5}{c}{$k_{\rm max}(z)$} \\
           & & $z=0$ & 0.3  & 0.7  & 1    & 1.5 \\
    Methods& & $D=1$ & 0.87 & 0.72 & 0.63 & 0.52 \\ \hline
    Linear & & 0.03& 0.08& 0.09& 0.09& 0.09  \\
1-loop SPT &  \cite{Peebles80,Juszkiewicz81,Vishniac83,Goroff86,Makino92,Jain94} & 
               0.08& 0.10& 0.11& 0.13& 0.14  \\
2-loop SPT & \cite{Fry94}& 
               0.04& 0.06& 0.09& 0.23& 0.20  \\
1-loop RPT & \cite{Crocce06a,Crocce06b,Crocce08}& 
               0.10& 0.13& 0.15& 0.16& 0.20  \\
2-loop RPT & & 
               0.08 &0.08& 0.11  & 0.11  & 0.13   \\
1-loop Closure & \cite{Taruya08}& 
               0.09& 0.10& 0.14& 0.15& 0.18  \\
2-loop Closure & & 
               0.08 & 0.13 & 0.17  & 0.27 & 0.21   \\
   Time-RG & \cite{Pietroni08}& 
               0.04& 0.05& 0.06& 0.09& 0.10  \\
   Large-N & \cite{Valageas07}& 
               0.08& 0.11& 0.12& 0.14& 0.17  \\
 Lag.Resum & \cite{Matsubara08}& 
               0.07& 0.08& 0.09& 0.10& 0.13  \\
   \hline
\end{tabular}
\caption{\label{tab:kmax}The methods we consider in this work and 
the lowest $k$ (in $h\,\text{Mpc}^{-1}$) at which each method departs
from our reference spectrum by 1\%, as a function of redshift for our chosen
$\Lambda$CDM cosmology.}
\end{table}

To gain an overview of the range of validity of the various methods we list in
Table \ref{tab:kmax} the smallest value of $k$ at which each method departs
from our reference spectrum by 1\% for $\Lambda$CDM (a comparison with other
schemes defined in the literature is presented in Table \ref{tab:converge}).
As expected, the methods perform better at smaller scales the higher the
redshift.
All of the methods out-perform linear theory, owing to the marked effects
of pre-virialization, however none of the methods appear to be accurate
beyond $k\simeq 0.1\,h\,\text{Mpc}^{-1}$ at $z=0$.

\subsection{Testing the dynamics}

While comparison of the power spectrum is the most common test for perturbation
theory, it is also useful to test if perturbation theory is correctly
describing the underlying dynamics.
To do so, we examine some of the constituent pieces from the simulations,
and compare to the perturbation theory predictions.

\begin{figure}
\resizebox{3.5in}{!}{\includegraphics{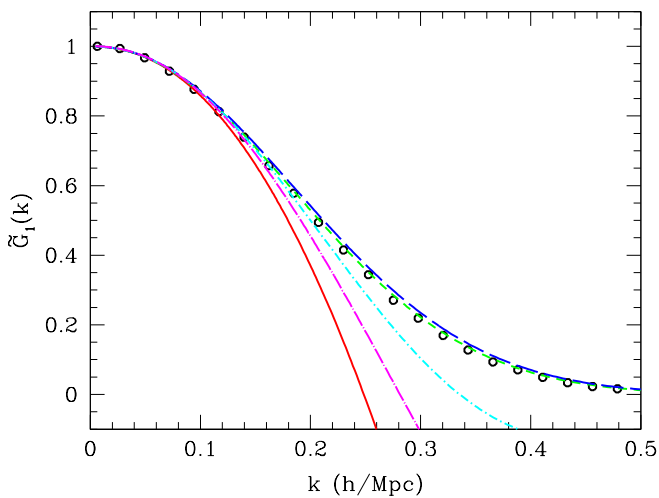}}
\caption{The non-linear propagator (normalized to 1 at $k=0$) for $\Lambda$CDM
at $z=0$.  The (red) solid line is SPT, the (green) short-dashed line is
Lagrangian resummation, the (blue) long-dashed line is RPT, the thick (cyan)
dot-short dashed line is closure theory and the thick (magenta) dot-long dashed
line is the large-$N$ expansion.}
\label{fig:propagator}
\end{figure}

Figure \ref{fig:propagator} compares the non-linear propagator
\begin{equation}
    \widetilde{G}_1(k) = G_{11}(k) + G_{12}(k)
        \sim \frac{\langle \delta_\text{NL}\delta_L^* \rangle}{\langle \delta_L\delta_L^* \rangle}
\end{equation}
from the simulations with the predictions of analytic models.
Only RPT and Lagrangian resummation give the expected behavior,
$\widetilde G_1 \to 0$, for large $k$.

\begin{table}
\begin{tabular}{c|c|cccc} \hline
Ref. & Method & \multicolumn{2}{c}{$\Lambda$CDM}
              & \multicolumn{2}{c}{$c$CDM}  \\
                   &     & $z=0$ & $z=1$ & $z=0$ & $z=1$ \\ \hline
\cite{Jeong06}     & SPT & 0.12  & 0.28  & 0.07  & 0.25 \\
\cite{Sefusatti07} & SPT & 0.10  & 0.19  & 0.08  & 0.19 \\
\cite{Nishimichi08}& SPT & 0.08  & 0.09  & 0.08  & 0.09 \\
\cite{Matsubara08} & Lag.R.& 0.08& 0.13  & 0.07  & 0.15 \\
\cite{Nishimichi08}& RPT & 0.09  & 0.09  & 0.09  & 0.10 \\
\cite{Nishimichi08}& Closure&0.09& 0.09  & 0.09  & 0.10 \\
\hline
\end{tabular}
\caption{\label{tab:converge}The value of $k$, in $h\,\text{Mpc}^{-1}$,
to which various flavors of perturbation theory can be trusted according
to various published criteria.  See \cite{Nishimichi08} for discussion.}
\end{table}

\begin{figure}
\resizebox{3.5in}{!}{\includegraphics{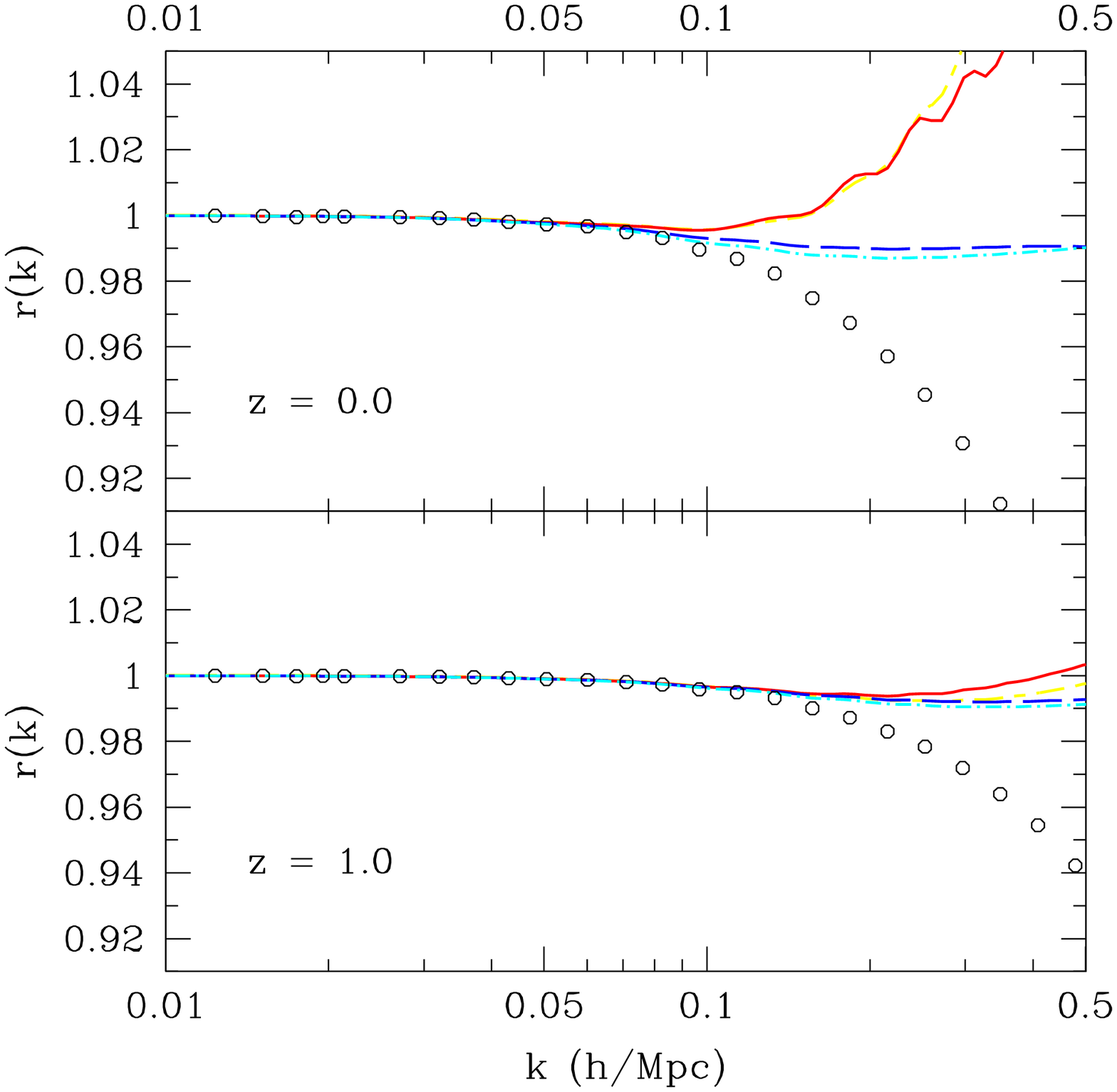}}
\caption{A comparison of the density-velocity cross-correlation predicted
analytically with that measured in simulations, for $\Lambda$CDM at redshift $z
= 0$ (top) and $z \approx 1$ (bottom).  As in Figure \ref{fig:previr}, the solid
(red) line is SPT, the dashed (blue) line is RPT, the dot-dashed (cyan) line is
closure theory, and the short-long-dashed (yellow) line is Time-RG theory.  For
simplicity we show only the 1-loop predictions for SPT, RPT, and closure
theory.}
\label{fig:rk}
\end{figure}

\begin{figure*}
\resizebox{3.5in}{!}{\includegraphics{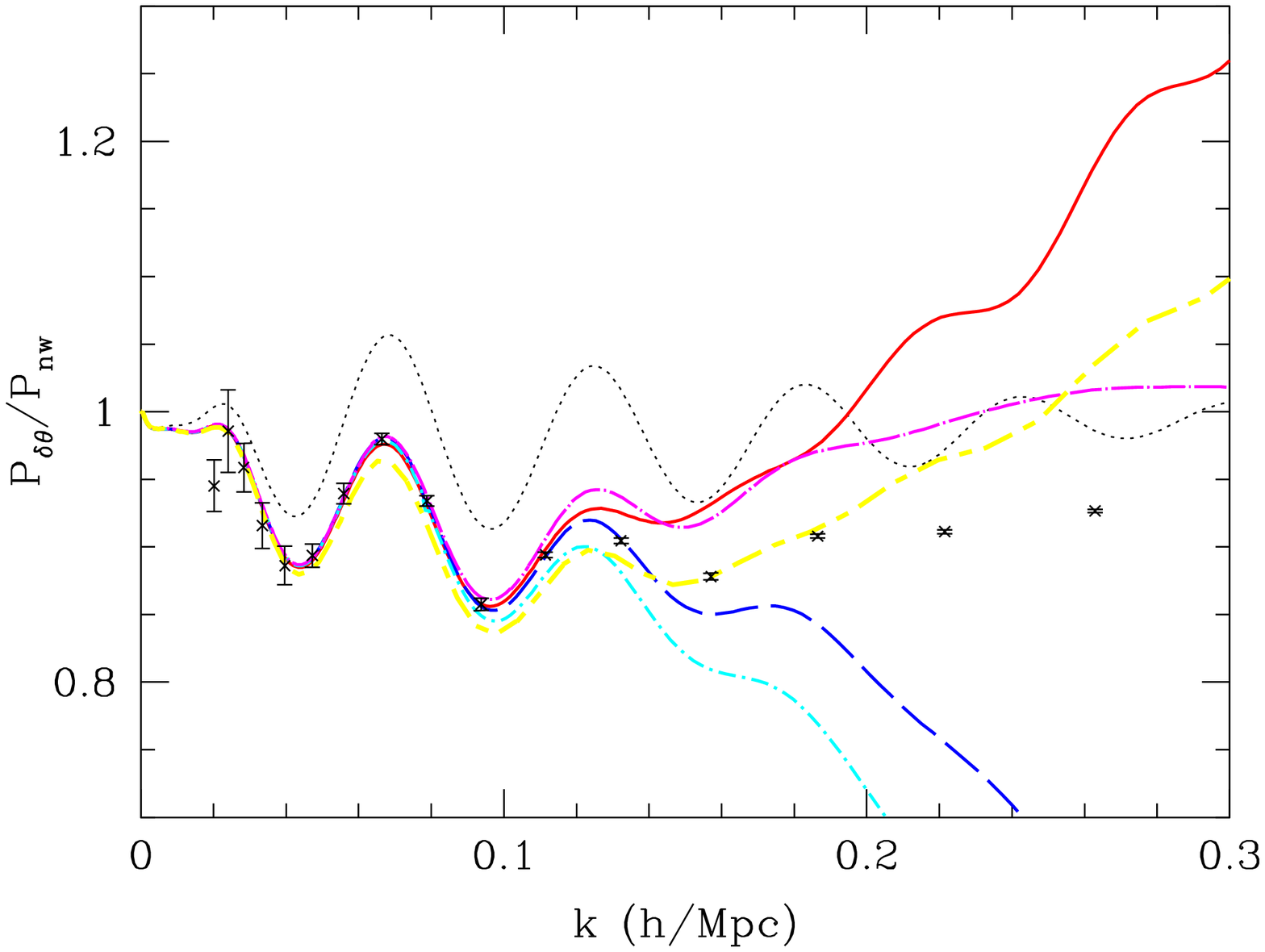}}
\resizebox{3.5in}{!}{\includegraphics{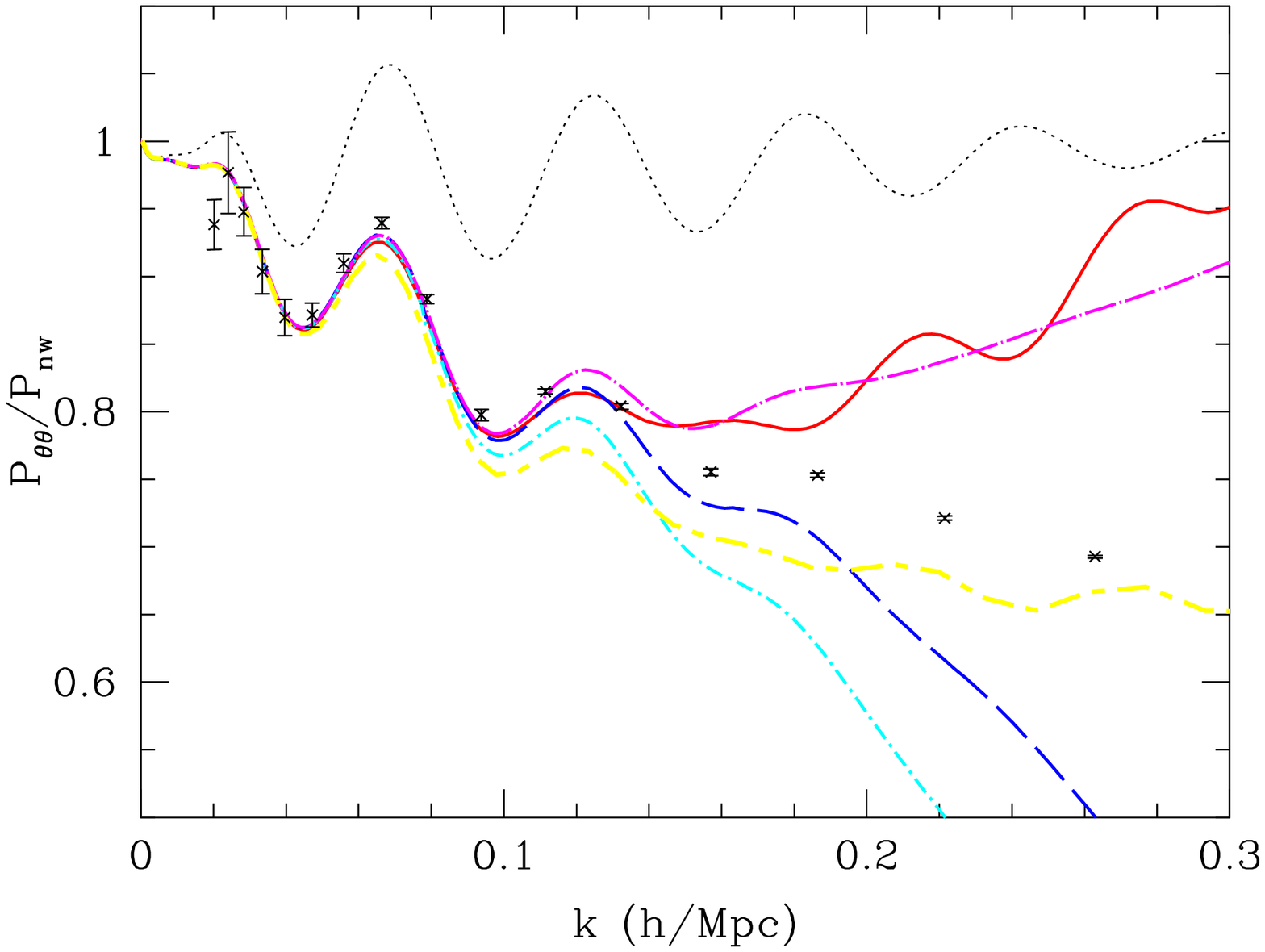}}
\caption{The density-velocity cross spectrum (left) and the velocity-velocity
auto-spectrum (right) for the $\Lambda$CDM model at $z=0$.  As in Figure
\protect\ref{fig:rk} the (black) dotted line is linear theory, the (red) solid
line is 1-loop SPT, the (blue) long-dashed line is 1-loop RPT, the (cyan)
dot-short dashed line is 1-loop closure theory, the thick (magenta) dot-long
dashed line is large-$N$ theory, and the thick (yellow) short-long dashed line
is time-RG theory.}
\label{fig:velocity}
\end{figure*}

Comparisons of perturbation theory with simulations typically focus on the
density auto-correlation function or power spectrum.  However perturbation
theory also makes predictions for the (irrotational) velocity field which can
be checked against simulations.  In Figure \ref{fig:rk} we show the
cross-correlation coefficient
\begin{equation}
  r(k) \equiv P_{\delta\theta}(k)/\sqrt{P_{\delta\delta}(k) P_{\theta\theta}(k)}
\end{equation}
for several theories, compared with the same quantity measured from simulations.
In linear theory $r(k) = 1$ identically.  On physical grounds one expects to
see a decoherence of density and velocity fields on small scales, and indeed
the simulations show $r(k) \to 0$ for large $k$.  None of the analytic theories
correctly reproduce this behavior.  SPT and Time-RG theory follow the downward
turn of the simulation data initially, but then predict an unphysical $r(k) >
1$ very soon after non-linear corrections become important.  RPT and closure
theory perform somewhat better, in that $r(k)$ never exceeds unity, but the
level of agreement with simulations is still not good above $k\simeq 0.1\,h
\text{ Mpc}^{-1}$.  (Note that we have displayed here only the 1-loop
predictions from these theories.)
The deviation in $r(k)$ seems to be driven mostly by the densities, with
perturbation theory performing better at the same scale for
the velocities than the densities (see Figure \ref{fig:velocity}).

\section{Discussion} \label{sec:discuss}

Standard perturbation theory has a simple and direct theoretical motivation,
and results in explicit integral expressions at any order.
If taken to infinite order, it provides an exact solution (though to an
idealized problem).
While standard perturbation theory works well at high redshift and large
scales, our results indicate that the standard expansion is badly behaved at
the redshifts and scales most accessible to observation, in
that higher order terms are comparable in magnitude to lower order terms.
Although one expects the expansion to converge if taken to sufficiently high
order, this comes at a great computational cost.  With advances in raw
computing power it may one day become possible to perform the calculation to
the requisite order, but in the near future this approach seems impracticable.

On the other hand, it should be emphasized that SPT performs rather well
at high redshifts, $z\gtrsim 1$.  Figure \ref{fig:spt} shows that 2-loop SPT
at $z=1$ agrees with simulations to 1\% out to $k = 0.2\,h\,\text{Mpc}^{-1}$
or beyond (where the simulations themselves become unreliable).  At these redshifts SPT
not only provides a reasonable theoretical prediction for the matter power
spectrum on observationally relevant scales, but also an estimate of the
theoretical uncertainty on this prediction.

RPT is essentially a rearrangement of the standard expansion, so like SPT it is
an exact solution if carried out to all orders.  While this rearrangement
appears to improve the convergence properties of the perturbation series, it
makes it unclear what small quantity (if any) we are actually expanding in.
Furthermore, RPT does not actually provide closed-form expressions for the
power spectrum, but rather integral relations where $P_{ab}$ is expressed in
terms of mode-coupling integrals of itself.  Thus in addition to truncating the
loop expansion at finite order, a fully consistent implementation of RPT
requires solving for $P_{ab}$ according to an iterative scheme, of which the
explicit expressions presented in \cite{Crocce08} represent only the first
step.  The error associated with this approximation has (to our knowledge) yet
to be quantified.

Closure theory derives from a very different perturbative scheme than RPT, yet
the results obtained are superficially quite similar.  There is no obvious way
to provide error estimates on the results of closure theory, however, as the
closure equations are obtained from heuristic approximations rather than a
systematic expansion.  Furthermore the propagator in closure theory shows
unrealistic oscillations for large $k$.  As mentioned previously, the closure
equations are solved approximately in \cite{Taruya08} by means of a Born-like
expansion.  Recently
\cite{Hiramatsu09} an attempt has been made to solve the closure equations
numerically without resort to such a Born-like expansion.  The resulting
predictions for the power spectrum appear to agree better with simulations than
the results presented here, although it is difficult to draw any firm
conclusions from the information provided.

Time-RG theory is based on a single well-defined approximation: the vanishing
of the trispectrum.  The validity of this approximation can easily be checked
in simulations, and in principle this could allow one to quantify the
theoretical uncertainty in the method.  As most easily seen in Figure
\ref{fig:pk_diff}, although time-RG theory follows the general trends of our
reference spectrum over a wider range than other methods, it comes up short by
1-2\% over the entire quasi-linear regime.  It also gives an unphysical
prediction for the density-velocity cross-correlation.

The large-$N$ expansion utilizes more sophisticated theoretical machinery
than other resummation techniques.  While the path-integral formalism
for computing clustering statistics is exact, the errors introduced by the
large-$N$ expansion are difficult to quantify, as `$N$' is a fictitious
parameter.  Although the large-$N$ expansion corresponds to an infinite partial
resummation of the standard perturbative expansion, from our results it seems
that this resummation offers little improvement over 1-loop SPT in the
quasi-linear regime.  The grossly unphysical behavior of the propagator in this
theory is likely responsible for this effect.  As mentioned previously, we have
focused attention on the steepest-descent method rather than the 2PI effective
action method of \cite{Valageas07}.  The latter method produces a more
reasonable propagator, and likely results in a better prediction for the power
spectrum, although at an increased computational cost.

Like SPT, the Lagrangian resummation prescription of \cite{Matsubara08} also
results in easy to compute, explicit integral expressions.  These are well
behaved at large $k$, allowing e.g.~$\xi(r)$ to be computed, and there are
natural extensions to redshift space and to halo bias \cite{Matsubara08b}.  For
the real-space mass power spectrum considered here it offers a marginal
improvement over 1-loop SPT for $k\Sigma<1/2$, although the damping prefactor
strongly overcompensates as one moves further into the quasi-linear regime.

Our results have interesting implications for generating a suite of
simulations aimed at constraining the matter power spectrum.  If we can
trust perturbative methods for $k\Sigma<x_c$, then we can focus the
computational resources on higher $k$.  Assuming Gaussian fluctuations,
obtaining 1\% accuracy in a bin $(k;\Delta k)$ requires $2\times 10^4$ modes.
There are $(kL_{\rm box})^3\,(\Delta k/k)/(2\pi^2)$ modes from a periodic box
of side length $L_{\rm box}$, so our 1\% constraint at $k\Sigma\simeq x_c$
translates into
\begin{eqnarray}
  & & L_{\rm box} \simeq  \frac{\Sigma}{x_c}
  \left(\frac{2\pi^2 N}{\Delta k/k}\right)^{1/3} \nonumber \\
  & &\approx 3{\rm Gpc}\ \left(\frac{0.5}{x_c}\right)
  \left(\frac{\Sigma}{10{\rm Mpc}}\right)
  \left(\frac{N}{2\times 10^4}\right)^{1/3}
  \left(\frac{0.1}{\Delta k/k}\right)^{1/3} \,\,
\end{eqnarray}
or an equivalent volume of smaller simulations.  This constraint is most
difficult to meet at $z=0$, since $\Sigma$ is larger and the simulations
must be evolved for longer.
As an example with the default parameters listed above we would require
27 simulations, each $1\,h^{-1}$Gpc on a side, to obtain percent level
constraints on the power spectrum of $\Lambda$CDM in a 10\% band near
$k\simeq 0.1\,h\,{\rm Mpc}^{-1}$ at $z=0$
but at $z=1$ we could trust perturbation theory at this scale and focus
the simulations on $k\simeq 0.15\,h\,{\rm Mpc}^{-1}$ where three times
fewer simulations of the same size are needed.

\begin{table}
\begin{tabular}{c|ccc}
 $k$ & $\quad \Delta^2_\text{lin}(k)\quad$ & $\quad \Delta^2_\text{ref}(k)\quad$ & $\quad G(k)\quad$ \\ \hline
0.02 &    0.012 & 0.012 & 0.996 \\
0.04 &    0.053 & 0.053 & 0.980 \\
0.06 &    0.130 & 0.129 & 0.950 \\
0.08 &    0.210 & 0.210 & 0.914 \\
0.10 &    0.274 & 0.285 & 0.859 \\
0.12 &    0.398 & 0.410 & 0.804 \\
0.14 &    0.466 & 0.507 & 0.737 \\
0.16 &    0.533 & 0.617 & 0.664 \\
0.18 &    0.662 & 0.764 & 0.592 \\
0.20 &    0.720 & 0.894 & 0.518 \\
\hline
\end{tabular}
\caption{Our input linear theory spectrum, at $z=0$, for the $\Lambda$CDM
model as a function of wavenumber (in $h\,{\rm Mpc}^{-1}$) and the reference
spectrum and propagator [$G(k)$] from our N-body simulations.
Our convergence tests indicate the spectra should be accurate to $<1\%$ over
the range of scales shown.}
\label{tab:reference}
\end{table}

\section{Conclusions} \label{sec:conclude}

Perturbative methods have a long history in cosmology, and are widely used
in many fields of physics.  Many of the techniques reviewed herein were first
developed in other fields and applied to other problems, with varying levels
of success, before being pressed into service for modeling cosmological
perturbations.
In this paper we have studied a variety of these methods as applied to
predicting the large-scale clustering of cold, collisionless matter in
an expanding Universe.
Our results indicate that the analytic theories correctly model the
approach to non-linearity and work well when the perturbations are small,
but none of the available theories are up to the challenge of fully
describing the behaviour of matter on quasi-linear scales at late times.
We have emphasized the need to study a range of different cosmologies and
to look at a variety of different statistical observables, as accidental
agreement between theory and simulations is possible if one only considers
the power spectrum.  We have computed the 2-loop contribution to SPT and
found that the standard perturbative expansion is badly behaved at low
redshifts, even on scales where 1-loop SPT was previously believed to be
valid.  This provides further motivation for studying
alternative analytic approaches based on non-perturbative methods, though at
the same time it emphasizes the need for error control in analytic methods.

This work has made use of a large number of high dynamic range N-body
simulations, against which we can compare the analytic models.  We make these
data public in Table \ref{tab:reference} to aid future work in the field.
In addition a flexible software package that implements the perturbation
schemes described in this paper is available from the authors.

\section*{Acknowledgments}

We thank Rom\'{a}n Scoccimarro, Mart\'{i}n Crocce, Pat McDonald, Salman Habib,
and Katrin Heitmann for helpful discussions.  The simulations presented in this
paper were carried out using computing resources of the National Energy
Research Scientific Computing Center and the Laboratory Research Computing
project at Lawrence Berkeley National Laboratory.  NP is supported by NASA
Hubble Fellowship NASA HST-HF-01200.01 and an LBNL Chamberlain Fellowship.

\vskip 12pt
\hrule

\appendix
\onecolumngrid

%

\section{Eulerian Perturbation Theory} \label{sec:eptreview}

Here we briefly recap the derivation of the fluid equations in the Eulerian
picture, and the assumptions that are made in perturbative treatments
(\cite{Peebles80,Juszkiewicz81,Vishniac83,Goroff86,Makino92,Jain94}; see
\cite{Bernardeau02} for a review).  The matter content of the Universe is
modeled as a large collection of identical particles of mass $m$, interacting
only through mutual gravitational attraction.  For low densities and
sub-horizon scales, such forces are adequately described by Newtonian gravity
in a uniformly expanding background, with the Newtonian potential sourced by
inhomogeneities in the density field.  The distribution function for such a set
of particles obeys the Vlasov equation.  The N-body methods are essentially a
Monte-Carlo evolution of the Vlasov equation where the Monte-Carlo tracer
super-particles move along characteristics.

Analytically one typically invokes the \emph{single-stream approximation},
which assumes that all particles at a given point $\vec{x}$ move together with
the same velocity $\vec{v}(\vec{x})$.  This amounts to demanding that
$f(\vec{x},\vec{p}) \propto \delta_D[\vec{p}-ma\vec{v}(\vec{x})]$, where $f$ is
the distribution function and $\delta_D$ is the Dirac delta function.  This
assumption is explicitly violated once shell crossing occurs in gravitational
collapse, but is thought to be a reasonable approximation for small density
constrasts.  The velocity moments of the Vlasov equation then give the familiar
fluid equations (e.g.\,\cite{Peebles80})
\begin{align}
    \label{eq:fluid}
    \frac{\partial\delta}{\partial\tau} + \grad\cdot[(1+\delta)\vec{v}] &= 0, \\
    \frac{\partial\vec{v}}{\partial\tau} + \mathcal{H}\vec{v} +
    (\vec{v}\cdot\grad)\vec{v} + \grad\Phi &= 0.
\end{align}
where $\mathcal{H} = d\ln a/d\tau = aH$ is the conformal Hubble parameter.

It is conventional to further assume that the vorticity
$\vec{w} = \grad\times\vec{v}$ of the velocity field vanishes, i.e.~that the
fluid is \emph{irrotational}.  This assumption is motivated by noting that
$\vec{w} \propto a^{-1}$ at linear order, and is well supported by simulations
\cite{PercivalWhite08,Pueblas08}.  Under this approximation the velocity field
is completely specified by its divergence $\theta = \grad\cdot\vec{v}$, and the
fluid equations reduce to
\begin{align}
    \frac{\partial\delta}{\partial\tau} + \theta
        &= -\grad\cdot(\delta\vec{v}), \\
    \frac{\partial\theta}{\partial\tau} + \mathcal{H}\theta + 4\pi Ga^2\rhobar\delta
        &= -\grad\cdot[(\vec{v}\cdot\grad)\vec{v}].
\end{align}
In Fourier space ${\vec{v}}(\vec{k})=-i\vec{k}\theta(\vec{k})/k^2$, giving
\begin{align}
    & \frac{\partial\delta(\vec{k})}{\partial\tau} + \theta(\vec{k})
    = -\int \frac{d^3q_1 d^3q_2}{(2\pi)^3} \delta_D(\vec{q}_1+\vec{q}_2-\vec{k})
    \frac{\vec{k}\cdot\vec{q}_1}{q_1^2}\theta(\vec{q}_1)\delta(\vec{q}_2), \\
    & \frac{\partial\theta(\vec{k})}{\partial\tau} +
    \mathcal{H}\theta(\vec{k}) + \frac{3}{2}\Omega_m \mathcal{H}^2
    \delta(\vec{k})
   = -\int \frac{d^3q_1 d^3q_2}{(2\pi)^3} \delta_D(\vec{q}_1+\vec{q}_2-\vec{k})
  \frac{k^2 (\vec{q}_1\cdot\vec{q}_2)}{2q_1^2 q_2^2}
  \theta(\vec{q}_1)\theta(\vec{q}_2).
\end{align}

As long as $\delta$ and $\theta$ are small, the right-hand sides of the fluid
equations are small and can be dropped; this approximation defines linear
theory.  The solution to the resulting linearized fluid equations may be
written as
\begin{equation}
    \label{eq:linsoln}
    \delta_L(\vec{k};z) = \frac{D(z)}{D(z_i)} \delta_i(\vec{k}) \quad , \quad
    \theta_L(\vec{k};z) = -\mathcal{H}(z)f(z)\frac{D(z)}{D(z_i)}\delta_i(\vec{k}),
\end{equation}
where $\delta_i$ is the density contrast at some early time $z_i$ when
linear theory is certainly valid, $D$ is the linear growth function (normalized
to 1 today), and $f \equiv d\ln D/d\ln a$.  At early times $\Omega_m \approx 1$
and $D \propto a$.  Note that a possible decaying mode contribution,
proportional to $a^{-3/2}$ at early times, is forced to zero in linear theory
by the condition that $\delta$ be well-behaved as $a \to 0$.  Note also that
the mode-coupling integrals vanish for $\vec{k}=0$, so linear theory
is always valid in some neighborhood of $\vec{k}=0$, even at late times.  For
convenience we define $\delta_0$ to be the linear density contrast today,
i.e.~$\delta_0(\vec{k}) = \delta_L(\vec{k};z=0)$.

It often proves convenient to use $\eta = \ln D$ as a time variable, and to
combine $\delta$ and $\theta$ into a two-component field
\begin{equation}
    \label{eq:fields}
    \Phi_a(\vec{k}) = \pmat{\delta(\vec{k}) \\ -\theta(\vec{k})/\mathcal{H}f}.
\end{equation}
If we introduce
\begin{equation}
    \alpha(\vec{q}_1,\vec{q}_2) = \frac{\vec{k}\cdot\vec{q}_1}{q_1^2} \quad,\quad
    \beta(\vec{q}_1,\vec{q}_2) = \frac{k^2 (\vec{q}_1\cdot\vec{q}_2)}{2q_1^2 q_2^2}
\end{equation}
the fluid equations may be recast as
\begin{equation}
  \label{eq:fluidphi}
  \left[\delta_{ab} \frac{\partial}{\partial\eta} + \Omega_{ab}\right]
  \Phi_b(\vec{k};\eta)
  = \int \frac{d^3q}{(2\pi)^3}\ \gamma_{abc}(\vec{q},\vec{k}-\vec{q})
  \Phi_b(\vec{q};\eta) \Phi_c(\vec{k}-\vec{q};\eta),
\end{equation}
where
\begin{equation}
    \Omega_{ab}(\eta)
        = \pmat{0 & -1 \\ -\frac{3\Omega_m}{2f^2} & \frac{3\Omega_m}{2f^2}-1}
\end{equation}
and the vertex $\gamma_{abc}(\vec{q}_1,\vec{q}_2)$ only has nonzero entries
$\gamma_{121}(\vec{q}_1,\vec{q}_2)=\gamma_{112}(\vec{q}_2,\vec{q}_1)=\alpha(\vec{q}_1,\vec{q}_2)/2$
and $\gamma_{222}(\vec{q}_1,\vec{q}_2)=\beta(\vec{q}_1,\vec{q}_2)$.
The initial fields at time $\eta_i$ are denoted
\begin{equation}
    \phi_a(\vec{k}) \equiv \Phi_a(\vec{k};\eta_i) = \delta_i(\vec{k})\pmat{1 \\ 1},
\end{equation}
and the linear theory solution is simply
$\Phi_a^{(L)}(\vec{k};\eta) = e^{\eta-\eta_i} \phi_a(\vec{k})$.

\subsection{Beyond linear order}

Standard perturbation theory (hereafter SPT;
\cite{Peebles80,Juszkiewicz81,Vishniac83,Goroff86,Makino92,Jain94})
defines a systematic series solution to the fluid equations \eqref{eq:fluidk}
in powers of the initial density contrast $\delta_i$ (or equivalently in
powers of the current linearly evolved density contrast $\delta_0$).
In an Einstein-de Sitter universe, where $\mathcal{H} \propto a^{-1/2}$ and 
$\Omega_m \mathcal{H}^2 \propto a^{-1}$, the expansion may be written as
\begin{equation}
    \delta(\vec{k};\tau) = \sum_{n=1}^\infty a^n(\tau) \delta_n(\vec{k}), \qquad
    \theta(\vec{k};\tau) = -\mathcal{H}(\tau) \sum_{n=1}^\infty a^n(\tau) \theta_n(\vec{k}),
\end{equation}
where $\delta_n(\vec{k})$ and $\theta_n(\vec{k})$ are time-indepedent
mode-coupling integrals over $n$ powers of the initial density field:
\begin{equation}
    \label{eq:deltan}
    \pmat{\delta_n(\vec{k}) \\ \theta_n(\vec{k})}
        = \int \frac{d^3q_1 \dots d^3q_n}{(2\pi)^{3n}} (2\pi)^3
        \delta_D\left(\sum\vec{q}_i-\vec{k}\right)
        \pmat{F_n(\{\vec{q}_i\}) \\ G_n(\{\vec{q}_i\})}
        \delta_0(\vec{q}_1) \dots \delta_0(\vec{q}_n).
\end{equation}
The kernels $F_n$ and $G_n$ satisfy recurrence relations that follow
straightforwardly from the equations of motion \cite{Goroff86,Makino92,Jain94}:
\begin{align}
    \label{eq:Fn}
    F_n(\vec{q}_1,\dots,\vec{q}_n)
        &= \sum_{m=1}^{n-1} \frac{G_m(\vec{q}_1,\dots,\vec{q}_m)}{(2n+3)(n-1)}
            \Big[(1 + 2n) \frac{\vec{k}\cdot\vec{k}_1}{k_1^2} F_{n-m}(\vec{q}_{m+1},\dots,\vec{q}_n)
               + \frac{k^2(\vec{k}_1\cdot\vec{k}_2)}{k_1^2 k_2^2} G_{n-m}(\vec{q}_{m+1},\dots,\vec{q}_n) \Big], \\
    \label{eq:Gn}
    G_n(\vec{q}_1,\dots,\vec{q}_n)
        &= \sum_{m=1}^{n-1} \frac{G_m(\vec{q}_1,\dots,\vec{q}_m)}{(2n+3)(n-1)}
            \Big[3 \frac{\vec{k}\cdot\vec{k}_1}{k_1^2} F_{n-m}(\vec{q}_{m+1},\dots,\vec{q}_n)
               + n \frac{k^2(\vec{k}_1\cdot\vec{k}_2)}{k_1^2 k_2^2} G_{n-m}(\vec{q}_{m+1},\dots,\vec{q}_n) \Big],
\end{align}
where $\vec{k}_1 = \vec{q}_1 + \dots + \vec{q}_m$, $\vec{k}_2 = \vec{q}_{m+1} +
\dots + \vec{q}_n$, $\vec{k} = \vec{k}_1 + \vec{k}_2$ and $F_1 = G_1 = 1$.

While the Einstein-de Sitter approximation is convenient, it is not 
necessary \cite{Takahashi08}. However we have confirmed that an accurate
approximation is to substitute the growth factor $D(z)$ for $a$,
\begin{equation}
    \delta(\vec{k};z) = \sum_{n=1}^\infty D^n(z) \delta_n(\vec{k}), \qquad
    \theta(\vec{k};z) = -\mathcal{H} f \sum_{n=1}^\infty D^n(z) \theta_n(\vec{k}),
\end{equation}
with the same mode-coupling integrals as above for $\delta_n$ and $\theta_n$.
The validity of this approximation is ultimately traced to the
fact that the ratio $\Omega_m/f^2$ is very nearly unity over the entire
lifetime of the universe for $\Lambda$CDM cosmologies, since $f \approx
\Omega_m^{0.6}$ \cite{Bernardeau02}.

\begin{figure}
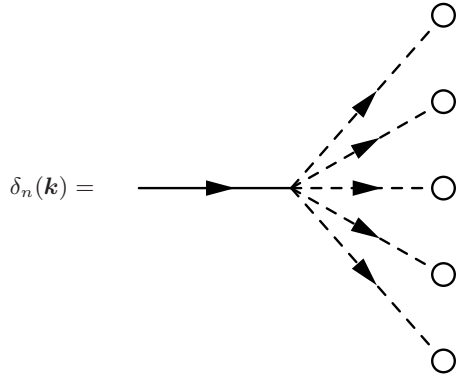

\begin{equation*}
    \delta_n(\vec{k}) =
    \fmfbox{55mm}{45,50}{0,0}{
        \fmfleft{a}
        \fmfrightn{r}{5}
        \fmfforce{(0.9w,0.04h)}{r1}
        \fmfforce{(0.9w,0.27h)}{r2}
        \fmfforce{(0.9w,0.5h)}{r3}
        \fmfforce{(0.9w,0.73h)}{r4}
        \fmfforce{(0.9w,0.96h)}{r5}
        \fmf{plain_arrow,label=$\vec{k}$,label.side=left,tension=5}{a,v}
        \fmf{dashes_arrow,label=$\vec{q}_n$}{v,r1}
        \fmf{dashes_arrow,label=}{v,r2}
        \fmf{dashes_arrow,label=}{v,r3}
        \fmf{dashes_arrow,label=}{v,r4}
        \fmf{dashes_arrow,label=$\vec{q}_1$,label.side=left}{v,r5}
        \fmfv{label=$F_n$,label.angle=120,label.dist=0.02w}{v}
        \fmfv{label=$\delta_0(\vec{q}_n)$,label.angle=0}{r1}
        \fmfv{label=...,label.angle=0}{r3}
        \fmfv{label=$\delta_0(\vec{q}_2)$,label.angle=0}{r4}
        \fmfv{label=$\delta_0(\vec{q}_1)$,label.angle=0}{r5}
        \fmfo{r1,r2,r3,r4,r5}
    }
\end{equation*}
\caption{Diagrammatic representation of the $n^\text{th}$ order contribution
to $\delta(\vec{k})$.}
\label{fig:fielddiag}
\end{figure}

To compute statistical observables it is convenient to introduce diagrammatic
rules for keeping track of the various terms in the perturbation series
\cite{Goroff86}.  The function $\delta_n(\vec{k})$ (or $\theta_n(\vec{k})$) may
be represented as in Figure \ref{fig:fielddiag}, where the open circles denote
factors of $\delta_0$, and the vertex denotes a momentum-conserving integral of
$F_n$ (or $G_n$) over intermediate wavevectors $\vec{q}_i$.  Algebraically the
$n^\text{th}$ order contribution $P^{(n)}$ is obtained by isolating all terms
of order $(\delta_0)^{2n}$ from the ensemble average
$\langle \delta(\vec{k}) \delta(\vec{k}') \rangle$, i.e.\
\begin{equation}
    (2\pi)^3 \delta_D(\vec{k}+\vec{k}') P^{(n)}(k;z)
        = D^{2n}(z) \sum_{m=1}^{2n-1} \langle \delta_m(\vec{k}) \delta_{2n-m}(\vec{k}') \rangle.
\end{equation}
The quantity $\langle \delta_m(\vec{k}) \delta_{2n-m}(\vec{k}') \rangle$ may be
represented diagrammatically by ``multiplying'' the diagrams for
$\delta_m(\vec{k})$ and $\delta_{2n-m}(\vec{k}')$.  Since the initial field
$\delta_i$ (and hence $\delta_0$) is Gaussian, ensemble averages of powers of
$\delta_0$ may be expanded in terms of the 2-point function $P_0$ according to
Wick's theorem.  Then the product of the diagrams $\delta_m(\vec{k})$ and
$\delta_{2n-m}(\vec{k}')$ is given by summing over all possible pairings of
their open circles, where open circles are paired according to the rule
\begin{equation}
    \fmfbox{18mm}{18,5}{0,0}{
        \fmfleft{il}
        \fmfright{v}
        \fmf{dashes_arrow,label=$\vec{q}$ ,label.side=left}{il,v}
        \fmfo{v}
    }
    \quad \times \quad
    \fmfbox{18mm}{18,5}{0,0}{
        \fmfleft{v}
        \fmfright{ir}
        \fmf{dashes_arrow,label=$\vec{q}'$,label.side=right}{ir,v}
        \fmfo{v}
    }
    \quad = \quad
    \fmfbox{35mm}{35,5}{0,0}{
        \fmfleft{il}
        \fmfright{ir}
        \fmf{dashes_arrow,label=$\vec{q}$ ,label.side=left}{il,v}
        \fmf{dashes_arrow,label=$\vec{q}'$,label.side=right}{ir,v}
        \fmfocross{v}
    }
    \quad \equiv \quad (2\pi)^3 \delta_D(\vec{q}+\vec{q}') P_0(q),
\end{equation}
with the additional understanding that any diagram containing a tadpole (a
fragment connected to the rest of the diagram by a single edge) vanishes
identically.

\begin{figure}
\begin{align*}
    \fmfbox{30mm}{30,30}{0,0}{
        \fmfleft{il}
        \fmfright{r1,r2}
        \fmfforce{(0.9w,0.1h)}{r1}
        \fmfforce{(0.9w,0.9h)}{r2}
        \fmf{plain_arrow,tension=2}{il,v}
        \fmf{dashes_arrow,label=}{v,r1}
        \fmf{dashes_arrow,label=}{v,r2}
        \fmfo{r1,r2}
    }
    \times
    \fmfbox{30mm}{30,30}{0,0}{
        \fmfleft{l1,l2}
        \fmfright{ir}
        \fmfforce{(0.1w,0.1h)}{l1}
        \fmfforce{(0.1w,0.9h)}{l2}
        \fmf{plain_arrow,tension=2}{ir,v}
        \fmf{dashes_arrow,label=}{v,l1}
        \fmf{dashes_arrow,label=}{v,l2}
        \fmfo{l1,l2}
    }
    \quad &= \quad 2 \;
    \fmfbox{50mm}{50,30}{0,0}{
        \fmfleft{il}
        \fmfright{ir}
        \fmf{plain_arrow,label=$\vec{k}$,label.side=left,tension=2}{il,v1}
        \fmf{plain_arrow,label=-$\vec{k}$,tension=2}{ir,v2}
        \fmf{dashes_arrow,label=$\vec{q}$}{v1,p1}
        \fmf{dashes_arrow,label=$\vec{k}-\vec{q}$}{v1,p2}
        \fmf{dashes_arrow,label=$-\vec{q}$}{v2,p1}
        \fmf{dashes_arrow,label=$\vec{q}-\vec{k}$}{v2,p2}
        \fmfforce{(0.5w,0.9h)}{p1}
        \fmfforce{(0.5w,0.1h)}{p2}
        \fmfocross{p1,p2}
    } \\
    &= 2 \int \frac{d^3q}{(2\pi)^3} F_2(\vec{q},\vec{k}-\vec{q}) F_2(-\vec{q},\vec{q}-\vec{k}) P_0(q) P_0(|\vec{k}-\vec{q}|)
\end{align*}
\caption{Diagrammatic prescription for computing $P^{(2,2)}(k)$.  The overall
factor of 2 comes from the two equivalent ways of pairing the open circles.
Only the single wavevector $\vec{q}$ must be integrated over, the rest being
determined by momentum conservation at vertices and translational invariance of
the 2-point function.}
\label{fig:diag22}
\end{figure}

As an example we show in Figure \ref{fig:diag22} how to obtain the 2nd order
contribution $P^{(2,2)}(k)$.  Notice that after invoking momentum conservation
at vertices and translational invariance of the 2-point function, only a single
wavevector remains to be integrated.  In general all diagrams contributing to
$P^{(n)}$ contain $n-1$ loops, requiring integration over $n-1$ independent
wavevectors.  For this reason we often classify power spectrum terms by their
number of loops rather than their ``order,'' which is a potentially ambiguous
concept.

With this expansion, statistical observables may be computed straightforwardly
in SPT to any fixed order.  For example, the first correction to the matter
power spectrum (second order in the initial power spectrum, fourth order
in the initial density contrast, or 1-loop in the diagrammatic idiom)
is given by
\begin{equation}
    \label{eqn:pk_spt}
    P(k) = P_L(k) + P^{(2,2)}(k) + P^{(1,3)}(k)
\end{equation}
where $P_L(k;z) = D^2(z) P_0(k)$ is the linear power spectrum and
\cite{Makino92}
\begin{subequations}
    \label{eqn:spt2}
\begin{align}
    \label{eqn:p13}
    P^{(1,3)}(k) &=
        \frac{1}{252} \frac{k^3}{4\pi^2} P_L(k) \int_0^\infty dr~ P_L(kr)
        \left[\frac{12}{r^2} - 158 + 100r^2 - 42r^4
            + \frac{3}{r^2}(r^2-1)^3 (7r^2+2) \ln\left|\frac{1+r}{1-r}\right| \right], \\
    \label{eqn:p22}
    P^{(2,2)}(k) &=
        \frac{1}{98} \frac{k^3}{4\pi^2} \int_0^\infty dr~ P_L(kr) \int_{-1}^1 dx~
        P_L\left(k\sqrt{1+r^2-2rx}\right) \frac{(3r+7x-10rx^2)^2}{(1+r^2-2rx)^2}.
\end{align}
\end{subequations}
At low $k$, $P^{(2,2)}$ is positive while $P^{(1,3)}$ is negative, and there is a
large degree of cancellation between them.  For large $k$
\begin{equation}
    P^{(2,2)}(k) \sim \frac{1}{4} \Sigma^2 k^2 P_L(k) \quad \text{and} \quad
    P^{(1,3)}(k) \sim -\frac{1}{2} \Sigma^2 k^2 P_L(k)
\end{equation}
where $\Sigma$ is defined by Eq.~\eqref{eqn:nlscale}, so for sufficiently large
$k$ the total second order contribution is negative.

It is also straightforward to derive expressions for the velocity power
spectrum \cite{Makino92}
\begin{align}
    P^{(1,3)}_{\theta\theta}(k) &=
        \frac{1}{84} \frac{k^3}{4\pi^2} P_L(k) \int_0^\infty dr~ P_L(kr)
        \left[\frac{12}{r^2} - 82 + 4r^2 - 6r^4
            + \frac{3}{r^2}(r^2-1)^3 (r^2+2) \ln\left|\frac{1+r}{1-r}\right| \right], \\
    P^{(2,2)}_{\theta\theta}(k) &=
        \frac{1}{98} \frac{k^3}{4\pi^2} \int_0^\infty dr~ P_L(kr) \int_{-1}^1 dx~
        P_L\left(k\sqrt{1+r^2-2rx}\right) \frac{(7x-r-6rx^2)^2}{(1+r^2-2rx)^2}.
\end{align}
the density-velocity cross-spectrum (e.g.~\cite{Scoccimarro04})
\begin{align}
    P^{(1,3)}_{\delta\theta}(k) &=
        \frac{1}{252} \frac{k^3}{4\pi^2} P_L(k) \int_0^\infty dr~ P_L(kr)
        \left[\frac{24}{r^2} - 202 + 56r^2 - 30r^4
        + \frac{3}{r^2}(r^2-1)^3 (5r^2+4) \ln\left|\frac{1+r}{1-r}\right| \right], \\
    P^{(2,2)}_{\delta\theta}(k) &=
        \frac{1}{98} \frac{k^3}{4\pi^2} \int_0^\infty dr~ P_L(kr) \int_{-1}^1 dx~
        P_L\left(k\sqrt{1+r^2-2rx}\right) \frac{(3r+7x-10rx^2)(7x-r-6rx^2)}{(1+r^2-2rx)^2},
\end{align}
and the propagator
\begin{equation}
  G(k) \simeq 1 + \frac{P^{(1,3)} + P^{(1,5)} + \cdots}{2\,P_L}
  \qquad .
\end{equation}

Though it is not usually considered, there is no real obstacle in going to
the next order in the systematic perturbative expansion described above.
For the third order (2-loop) contribution one finds
$P^{(3)}(k) = P^{(1,5)}(k) + P^{(2,4)}(k) + P^{(3,3)}(k)$ with \cite{Fry94}
\begin{align}
    P^{(1,5)}(k)
        &= 30 P_L(k) \int \frac{d^3q}{(2\pi)^3} \frac{d^3p}{(2\pi)^3} F_5^{(s)}(\vec{k},\vec{q},-\vec{q},\vec{p},-\vec{p}) P_L(q) P_L(p) \\
    P^{(2,4)}(k)
        &= 24 \int \frac{d^3q}{(2\pi)^3} \frac{d^3p}{(2\pi)^3} F_2^{(s)}(\vec{q},\vec{k}-\vec{q}) F_4^{(s)}(-\vec{q},\vec{q}-\vec{k},\vec{p},-\vec{p}) P_L(q) P_L(p) P_L(|\vec{k}-\vec{q}|) \\
    P^{(3,3)}(k)
        &= \int \frac{d^3q}{(2\pi)^3} \frac{d^3p}{(2\pi)^3} \Big[ 9 F_3^{(s)}(\vec{q},-\vec{q},\vec{k}) F_3^{(s)}(-\vec{k},\vec{p},-\vec{p}) P_L(k) P_L(q) P_L(p) \notag \\
        &\quad\qquad\qquad\qquad + 6 F_3(\vec{q},\vec{p},\vec{k}-\vec{q}-\vec{p}) F_3(-\vec{q},-\vec{p},\vec{q}+\vec{p}-\vec{k}) P_L(q) P_L(p) P_L(|\vec{k}-\vec{q}-\vec{p}|) \Big]
\end{align}
and with $F_n^{(s)}$ given by Eq.~\eqref{eq:Fn} symmetrized over its $n$
arguments $\vec{q}_1,\dots,\vec{q}_n$.  Using rotational symmetry to eliminate
one azimuthal integration, the resulting expressions require 5-dimensional
mode-coupling integrals which are best performed using Monte Carlo methods.

Expressions for higher order contributions are not difficult to derive, but the
computational costs of evaluating them quickly spiral out of control.  In
general the $\ell$-loop contribution requires mode-coupling integrals of
dimension $3\ell$ ($3\ell-1$ after rotational symmetry), making 1-loop simple,
2-loop possible, and higher orders impracticable.

\section{Lagrangian Perturbation Theory} \label{sec:lptreview}

The Lagrangian description of structure formation
\cite{Buchert89,Moutarde91,Hivon95}
relates the current, or Eulerian, position of a mass element, $\bx$, to
its initial, or Lagrangian, position, $\bq$, through a displacement vector
field: $\bx = \bq + \bP(\bq)$.
(Note that $\bq$ is used as a position vector in the Lagrangian picture,
whereas the same symbol is used as a wavevector in the Eulerian picture.)
The displacements can be related to overdensities by \cite{TayHam96}
\begin{equation}
  \delta(\bx) = \int d^3q\ \delta_D(\bx-\bq-\bP)-1 \quad , \quad
  \delta(\bk) = \int d^3q\ e^{-i \bk\cdot \bq}
  \left(e^{-i \bk\cdot \bP(\bq)} - 1\right) \ .
\label{eqn:lptdensity}
\end{equation}
The displacements evolve according to
\begin{equation}
  \frac{d^2\bP}{dt^2} + 2 H \frac{d \bP}{dt} =
  -\nabla_x \phi\left[\bq+\bP(\bq)\right] \,\,,
\end{equation}
where here and only here $\phi$ is the gravitational potential.
Analogous to Eulerian perturbation theory, standard LPT expands the
displacement in powers of the linear density field with \cite{BouColHivJus95}
\begin{equation}
  \bP^{(n)}(\bk) = \frac{i}{n!} \int
  \prod_{i=1}^n \left[\frac{d^3k_i}{(2\pi)^3} \right]
  \ (2\pi)^3\delta_D\left(\sum_i \bk_i-\bk\right)
  \bL^{(n)}(\bk_1,\cdots,\bk_n,\bk)
  \delta_0(\bk_1)\cdots\delta_0(\bk_n) \; .
\end{equation}
and the $\bL^{(n)}$ have closed form expressions in terms of dot products of
wave vectors which can be generated by recurrence relations.
Expanding the exponential in Eq.~(\ref{eqn:lptdensity}) we obtain a
perturbative series for the overdensity,
$\delta = \delta^{(1)}+\delta^{(2)}+\cdots$
where, e.g.,
\begin{equation}
\delta^{(2)}(\bk) = \frac{1}{2} \int\frac{d^3k_1d^3k_2}{(2\pi)^3}
   \delta_D(\bk_1 + \bk_2 - \bk) \delta_0(\bk_1) \delta_0(\bk_2)
  \left[\bk \cdot \bL^{(2)}(\bk_1, \bk_2, \bk)
  +  \bk \cdot \bL^{(1)}(\bk_1) \bk \cdot \bL^{(1)}(\bk_2) \right]
\end{equation}
is second order in the linear density field $\delta_0$.

A similar expansion can be performed for the power spectrum, which from
Eq.~(\ref{eqn:lptdensity}) can be written
\begin{equation}
  P(k) = \int d^3q\ e^{-i\vec{k}\cdot\vec{q}}
  \left( \left\langle e^{-i\vec{k}\cdot\Delta\bP}\right\rangle-1\right)
\label{eqn:lptpower}
\end{equation}
where $\Delta\bP=\bP(\bq)-\bP(0)$ and we have used translational invariance.

Alternatively \cite{Matsubara08} suggested using the cumulant expansion
theorem for the exponential in Eq.~\eqref{eqn:lptpower} and using the
binomial theorem to expand the term $(\bk\cdot\Delta\bP)^N$.
One obtains two types of terms: those depending on $\bP$ at the same point
and those depending on $\bP$ at two different points.
Owing to statistical homogeneity the first type of term is independent of
position and can be factored out of the integral leaving \cite{Matsubara08}
\begin{equation}
    P(k) = \exp\left[ -2\sum_{n=1}^\infty (-1)^{n-1}
        \left\langle [\vec{k}\cdot\vec{\Psi}(0)]^{2n} \right\rangle  \right]
        \int d^3r ~ e^{i\vec{k}\cdot\vec{r}}
            \left\{ \exp\left[ \sum_{N=2}^\infty \frac{k_{i_1} \dots k_{i_N}}{N!}
                    B^{(N)}_{i_1 \dots i_N}(\vec{r})\right] - 1 \right\},
\end{equation}
where $k_{i_1} \dots k_{i_N} B^{(N)}_{i_1 \dots i_N}(\vec{r})$ is shorthand
for the second type of term.

In a traditional perturbative calculation one would expand this expression to a
fixed order in $\vec{\Psi}$; this approach indeed reproduces the SPT result to
2nd order.  However one might expect that the position-independent cumulant
factors are more important on large scales than the position-dependent ones,
suggesting that these factors should be left unexpanded in the exponential.
Using well-known previous results from LPT the first corrections to the power
spectrum are then \cite{Matsubara08}
\begin{equation}
    P(k) = e^{-(k\Sigma)^2/2}
           \left[ P_L(k) + P^{(2,2)}(k) + \widetilde P^{(1,3)}(k) \right],
\end{equation}
where $\Sigma$ is given by Eq.~\eqref{eqn:nlscale}, $P^{(2,2)}(k)$ is as in SPT
[Eq.~\eqref{eqn:p22}] and
\begin{equation}
    \widetilde P^{(1,3)}(k) = \frac{1}{252} \frac{k^3}{4\pi^2} P_L(k)
        \int_0^\infty dr ~ P_L(kr) \left[\frac{12}{r^2} + 10 + 100r^2 - 42r^4 +
        \frac{3}{r^3} (r^2-1)^3 (7r^2+2) \ln\left|\frac{1+r}{1-r}\right| \right].
\end{equation}
Notice that this differs from the SPT result [Eq.~\eqref{eqn:p13}] only by the
replacement $-158 \to 10$ in the brackets.  If the exponential prefactor is
expanded to first order in $P_L$ the SPT result is recovered exactly
\cite{Matsubara08}.  Also note that the first term $e^{-\Sigma^2 k^2/2} P_L(k)$
is identical to the tree-level RPT result in the large-$k$ limit.

\bibliography{ptcompare}
\bibliographystyle{apsrev}

\end{fmffile}
\end{document}